\documentclass{report}
\usepackage{amsmath,amssymb,amsfonts,amscd,theorem,graphics}
\usepackage[latin1]{inputenc}
\renewcommand{\arraystretch}{1.5}

\def\simarrow{\mathrel{\raise -0.5mm\hbox{$\sim$}}\hspace{-1.8mm}{\rightarrow} } 

\def\bsimarrow{\leftarrow\hspace{-0.7mm}\mathrel{\raise -0.5mm\hbox{$\backsim$}} }

\def\bt{\begin{tabular}}
\def\te{\end{tabular}}

\def\lettrine#1#2#3{\noindent\hangindent#1\hangafter-#2
\hskip-#1\smash{\hbox to #1{#3\hfill}}\ignorespaces}

\def\BM{\begin{pmatrix}}
\def\EM{\end{pmatrix}}

\def\ds{\displaystyle}

\def\d=f{\buildrel\hbox{\scriptsize d\'{e}f}\over \Longleftrightarrow}

\def\square{\hfill\hbox{\vrule height .9ex width .8ex depth -.1ex}}
\def\rit{\text{\it I\hskip -2pt  R}}

\def\Bd{{\text B}}

\def\Ed{{\text E}}

\def\Hs{{\cal H}}

\def\Ls{{\cal L}}

\def\be{\begin{equation}}
\def\ee{\end{equation}}
\def\beqn{\begin{eqnarray}}
\def\eeqn{\end{eqnarray}}
\def\nobeqn{\begin{eqnarray*}}
\def\noeeqn{\end{eqnarray*}}
\def\ba{\left(\begin{array}}
\def\ea{\end{array} \right) }

\def\bpr{\paragraph{Proof.}}

\def\epr{\square\vskip 6pt}
\def\eop{\hbox{\vrule height .9ex width .8ex depth -.1ex}}

\def\o{\overline}
\def\and{\; \mbox{and} \;}

\newcommand{\half}{\frac{1}{2}}

\def\hfl#1#2{\smash{\mathop{\hbox to 12mm{\rightarrowfill}}
\limits^{\scriptstyle #1}_{\scriptstyle #2}}}

\def\Be{\begin{enumerate}}
\def\Ee{\end{enumerate}}

\def\Bena{\begin{enumerate}
\def\labelenumi{\theenumi)}
\def\theenumi{\arabic{enumi}}
\def\labelenumii{\theenumii)}
\def\theenumii{\alph{enumii}}}

\def\Bean{\begin{enumerate}
\def\labelenumii{\theenumii)}
\def\theenumii{\arabic{enumii}}
\def\labelenumi{\theenumi)}
\def\theenumi{\alph{enumi}}}

\def\Bero{\begin{enumerate}
\def\labelenumii{\theenumii)}
\def\theenumii{\arabic{enumii}}
\def\labelenumi{(\theenumi)}
\def\theenumi{\roman{enumi}}}

\def\BeRo{\begin{enumerate}
\def\labelenumii{\theenumii)}
\def\theenumii{\arabic{enumii}}
\def\labelenumi{(\theenumi)}
\def\theenumi{\Roman{enumi}}}

\def\Bi{\vskip 11pt\begin{itemize}\itemsep=18pt}
\def\bi{\begin{itemize}}
\def\Ei{\end{itemize}\vskip 11pt}
\def\ei{\end{itemize}}

\def\Bd{\begin{description}}
\def\Ed{\end{description}}

\def\R{\right}
\def\L{\left}
\def\F{\frac}

\def\bbf{\bfseries\boldmath}

\def\Bi{\begin{itemize}}
\def\Ei{\end{itemize}}

\def\sum{\mathop{\Sigma}\limits}
\def\prod{\mathop{\Pi}\limits}
\def\bigoplus{\mathop{\oplus}\limits}

\def\bplus{\mathop{\boxplus}\limits}

\def\Aa{{\mathbb{A}\,}}

\def\CC{{\mathbb{C}\,}}
\def\NN{{\mathbb{N}\,}}

\def\Repsp{\operatorname{Repsp}}

\def\Eis{\operatorname{EIS}}
\def\Ellip{\operatorname{ELLIP}}
\def\TAN{\operatorname{TAN}}

\def\RxL{_{R\times L}}
\def\RxDL{_{R\times_D L}}

\def\GL{\operatorname{GL}}

\def\MM{{\mathbb{M}\,}}

\def\EE{E}

{\theorembodyfont{\rmfamily}
  }
{\theorembodyfont{\rmfamily}
  }

{\theorembodyfont{\rmfamily}
  }
{\theorembodyfont{\rmfamily}
  }

{\theorembodyfont{\rmfamily}
  }

\def\lr{left (resp. right) }

\begin{document}

\pagestyle{empty}

\null
 \vfill
\begin{center} 
{\LARGE Brane and string field structure of elementary particles\par}%
   \vskip 3em
   {\large C. Pierre
\\[5mm]
Institut de Mathématique pure et appliquée\\
Université de Louvain\\
Chemin du Cyclotron, 2\\
B-1348 Louvain-la-Neuve\\ Belgium\\
pierre@math.ucl.ac.be\par}

\end{center}

\vfill
\centerline{MSC (2000): 81T10, 81T30, 81T70, 11F70}\eject

\null\vfill\vfill

{\begin{abstract}{\noindent 
\\
The main relevant features of quantum (field) theories are examined in order to set up the physical and mathematical foundations of the algebraic  quantum theory.
\newline
It then appears that the two quantizations of QFT, as well as the attempt of unifying it with general relativity, lead us to consider that the internal structure of an elementary fermion must be twofold and composed of three embedded internal (bi)structures which are vacuum and mass (physical)   bosonic fields decomposing into packets of pairs of strings behaving like harmonic oscillators characterized by integers $\mu $ corresponding to normal modes at $\mu $ (algebraic) quanta.
} 

\vfill

\mbox{}\hfill\begin{minipage}{10cm}
``The mathematicians, who studied physics, fail because the actual physical situations in the real world are so complicated that it is necessary to have a much broader understanding of the equations''.\hfill \mbox{R.P. Feynman.}
\vskip 11pt

``I understand what an equation means if I have a way of figuring out the characteristics of its solution without actually solving it''.
\mbox{}\hfill P.A.M. Dirac.
\vskip 11pt

\mbox{} \hfill (From Feynman lectures on physics -- II)
\end{minipage}

\vfill\eject\end{abstract}}

\pagestyle{myheadings}



\newpage

\def\thepage{\arabic{page}}

\setcounter{page}{1}

\chapter{Introduction}

\thispagestyle{empty}

\addtocontents{toc}{\protect\thispagestyle{empty}}

In the paper ``Algebraic quantum theory'' \cite{Pie4}, noted ``AQT'', a new quantum field theory of strings was introduced in order to endow the elementary particles with an algebraic space-time structure constituting their own vacua from which their mass shells can be generated.  This allows to find a way out to the inextricable problem of unifying general relativity (noted ``GR'') with quantum field theory (noted ``QFT'') in the sense that the expanding space-time of GR becomes now spreaded out discretely at the Planck scale around ``organizing centers'' of the internal vacua of the elementary particles.  Note that these ``organizing centres'' refer to attractors from a dynamical point of view.
\vskip 11pt

The mathematical foundations of AQT were rather well developed in \cite{Pie4} and initiated in \cite{Pie1}, \cite{Pie2} and \cite{Pie3}.  They include essentially:
\Bi
\item the Langlands global program based on the (in)finite dimensional representations of the (ir)reducible bilinear algebraic semigroups over products, right by left, of completions of a numberfield of characteristic 0.
\item the versal deformations of degenerate singularities and their blowups \cite{A-G-L-V}.

\item the algebraic representations of von Neumann bialgebras set on  bilinear Hilbert spaces.
\Ei
\vskip 11pt

But, the connections between the structure of AQT and the main attainments of quantum (and classical) field theories and string theories were not clearly shown up in \cite{Pie4}: it is thus the aim of this paper to remedy this gap while pointing out the main physical advances of this new quantum string field theory as for example:
\Bi
\item a better understanding of the physical phenomena at the elementary particle level due to the action-reaction processes between left and right semiobjects which are generated mathematically by envisaging a bilinear (non commutative) framework.

\item a good reason to see in the internal vacua of the elementary particles a candidate for the dark energy.
\Ei
\vskip 11pt

What is particularly important is to relate the two quantizations of quantum (field) theories to the main concepts of AQT and to show that they imply the mathematical structure of AQT.
\vskip 11pt

In this perspective, the main concepts of relativistic quantum mechanics, (classical and) quantum field theories and string theories are examined in a critical way in chapter 2 so that the relevant features of these theories could be separated in order to set up the physical foundations of a quantum theory of structure of elementary particles.

It then appears that the two quantizations of quantum (field) theories lead to consider the following conceptual basis for a new quantum structure of elementary particles:
\Be
\item {\bf the first quantization\/} of (relativistic) quantum mechanics suggests that:
\Be
\item a mathematical structure be given to the quanta; under the circumstances, they become algebraic closed irreducible real subsets characterized by a Galois extension degree equal to $N$~.

\item a bialgebra of operators acting on bilinear Hilbert spaces of fields be introduced as being a von Neumann bialgebra.
\Ee
\vskip 11pt

\item {\bf the second quantization\/} of QFT and its unification with GR leads to envisage that:
\Be
\item every elementary fermion must be viewed as an elementary bisemifermion which (see proposition 2.7):
\Bi
\item is localized in an open  ball.
\item is given by the product of a left semifermion, localized in the upper half space, and of a right symmetric semifermion, localized in the lower half space in such a way that, under some external perturbation, this bisemifermion could be split, generating a pair of fermion-antifermion, of which fermion corresponds to the left semifermion and antifermion to the right semifermion; by this way, the right semifermion (~$\approx$ antisemifermion), projected onto the associated left semifermion, is hidden by the only observable (left) fermion.

\item is composed of three central diagonal embedded bistructures, which are its internal structural fields, in such a way that the two most internal bistructures, labeled ``~$ST$~'' and ``~$MG$~'', are its internal vacuum from which its mass shell bistructure ``~$M$~'' can be created.
\Ei

\item Each central diagonal bistructure is a (bilinear) field, direct sum of a time field and of a space field, in such a way that each field is composed of (the sum of) the set of packets of pairs of strings (or bistrings), behaving like harmonic oscillators and characterized by integers $\mu $ corresponding to normal modes at $\mu$ quanta.
\Ee
\Ee

The string fields, included into the corresponding brane fields \cite{Joh}, are proved, in chapter 3, to correspond to (bisemi)sheaves of $\CC$-valued differentiable bifunctions on the conjugacy class representatives of algebraic bilinear semigroups over the real ramified completions of number fields of  characteristic 0~.


Thereafter, the holomorphic and automorphic representations of these string fields are studied in the second part of chapter 3.

Finally, in chapter 4, the consideration of von Neumann bialgebras on these (bilinear) fields allows to define the states of the fermionic vacuum (operator valued) fields and the states of the corresponding mass (operator valued) fields generated from versal deformations and blowups of singularities on the vacuum fields.

In this context, it is shown how mass open bistrings can be created from the vacuum fields and annihilated.

The paper ends with a brief survey of interacting fields, which are gravitational and electromagnetic off-diagonal fields generated from the consideration of the completely reducible modular bilinear {\bf non-orthogonal\/} representation spaces of bilinear algebraic semigroups.

All developments of this paper refer to the preprint ``Algebraic quantum theory'' \cite{Pie4}.

\chapter{From quantum field theories to the concept of fields in AQT}

\thispagestyle{empty}

\section{Underlying bilinearity in classical mechanics}

Let $X$ denote the manifold of positions of $r$ material points and let $M=T^*(X)$ be the total space of its cotangent bundle taking into account the positions and momenta of these points.
\vskip 11pt

Classical mechanics deals with differentiable functions on $M$~, interpreted as a phase space at $r$ degrees of freedom.  Such a differentiable function, extensely used in classical dynamics, is the function of Lagrange $\Ls(q_1,\cdots,q_r;\dot q_1,\cdots,\dot q_r,t)=T-U$~, where $T$ is the kinetic energy and $U$ is the potential energy of the considered system.

(Classical) Dynamics starts then up with the least action principle stating that the 
integral $\int^{t_1}_{t_0}\Ls\ dt$ must be stationary for an infinitesimally small variation of the movement between the initial state at time $t=t_0$ and the final state at time $t=t_1$ \cite{Bro2}.

The functions on $C^\infty (M)$ constitute the algebra of observables in classical dynamics and the points of $M$ are in fact classical states.
\vskip 11pt

A Lie algebra structure on $C^\infty (M)$ is reached by considering on $M$ a symplectic form $w=\sum^r_{j=1}dq_j\wedge dp_j$~, where $q_j$ are local coordinates and $p_j$ are the corresponding momenta.

The Poisson bracket operation \cite{Duf}
\[ \{f,g\}=\sum_j \L( \F{\partial f}{\partial q_j} \ \F{\partial g}{\partial p_j} 
- \F{\partial f}{\partial p_j} \ \F{\partial g}{\partial q_j} \R)\]
for the functions $f$ and $g$ on the algebra $C^\infty (M)$ corresponds to the symplectic form $w$ and is a $\CC$-bilinear operation $(f,g)\to\{f,g\}$ \cite{Maz1} satisfying $\{f,g\}=0$ and the Jacobi identity.

A general Poisson bracket operation on  $C^\infty (M)$ has the form:
\[ \{f,g\}(x) = \sum^r_{i,j=1} \alpha ^{i,j}(x) \
\F{\partial f}{\partial x_i} \ \F{\partial g}{\partial x_j} \]
where $\alpha ^{i,j}(x) $ is a skew-symmetric bivector field \cite{Maz1}.

A Poisson manifold is a manifold $M$ with Poisson brackets on $C^\infty (M)$~.

A dynamics, resulting from the Poisson bracket $\{f,H\}$~, is obtained
if the function of Hamilton $H(q_1,\cdots,q_r,p_1,\cdots,p_r,t)$~, playing the role of energy, is introduced on $C^\infty (M)$~.  Indeed, let
\[ dH=-\sum_j\dot p_j\ dq_j+\sum_j\dot q_j\ dp_j\]
be its differential leading to the equations of Hamilton:
\[ \dot q_j = \F{\partial H}{\partial p_j} \;,\qquad
\dot p_j = \F{\partial H}{\partial q_j} \;\cdotp\]
Then, the total derivative with respect to $t$ of $f(q_1,\cdots,q_r,p_1,\cdots,p_r,t)\in C^\infty (M)$~, expressed according to:
\[ \F{df}{dt}=\F{\partial f}{\partial t}+\sum_j\ \L( \F{\partial f}{\partial q_j}\ \dot q_j
+ \F{\partial f}{\partial p_j}\ \dot p_j\R)\;, \]
becomes
\[ \F{df}{dt}=\F{\partial f}{\partial t}+\{f,H\}\]
if the Hamilton equations are taken into account.  And, if \quad $\F{df}{dt}=0$~, \quad $\F{\partial f}{\partial t}+\{f,H\}=0$ \quad is the equation of the dynamics written in function of the Poisson bracket $\{f,H\}$ taking into account the energy of the system.
\vskip 11pt

\section{First quantization in the wave quantum mechanics}

\Bean
\item The first quantization of quantum mechanics leads to the main {\bf following change}:

The ``classical mechanics'' algebra $C^\infty (M)=C^\infty (T^*(X))$ of observables, 
which are differentiable functions (for example, $\Ls$ or $H$~) on the phase space 
$M$~, is replaced by the ``quantum mechanics'' algebra of operators acting on a 
linear Hilbert space $\Hs$ of states or quantum observables, this algebra of 
operators being the von Neumann algebra $M(\Hs)$ in $\Hs$~.

In this context, the generalized coordinates $q_1,\cdots,q_r$ and $p_1,\cdots,p_r$ of the $r$ material points become, in the quantum language, operators $q_1,\cdots,q_r$ and $p_1\to \F{\hbar }i\ \F\partial {\partial q_1}$~, \ldots, $p_r\to \F{\hbar }i\ \F\partial {\partial q_r}$~, respectively according to the correspondence rule.

If these $r$ material points are immersed in a 3-dimensional space, the system has $k=3r$ degrees of freedom.  The operators have to obey the Heisenberg commutation relations $[q_j,p_j]=i\hbar$ where the Planck's constant $\hbar $ is supposed to introduce the quantum aspect of the theory \cite{Dir4}, \cite{Con}.

Let $H(q_1,\cdots,q_{3r},p_1,\cdots,p_{3r})$ be the Hamilton's function of our system of $r$ material points which are interpreted as particles in the quantum perspective.

Quantum mechanics, following classical mechanics, tries to get from $H$ the energy levels of the system.
\vskip 11pt

\item {\bf Matrix aspect}

The procedure consists in finding a matricial representation to the operators 
$q_1,\cdots,q_{3r},p_1,\cdots,p_{3r}$ in such a way that the matrix
\[ W=H(Q_1,\cdots,Q_{3r},P_1,\cdots,P_{3r})\]
can be reduced to a diagonal matrix.

$Q_1,\cdots,Q_{3r}$ and $P_1,\cdots,P_{3r}$ are the matricial representations of
$q_1,\cdots,q_{3r}$ and $p_1,\cdots,p_{3r}$ satisfying the matrix commutation relations of  Heisenberg: this is the philosophy of the theory of matrices whose key papers can be found in \cite{Vdw}.

What is important to remark is that:
\Be
\item the rank(s) of these matrices $Q_1,\cdots,Q_{3r}$ and $P_1,\cdots,P_{3r}$ is (are) the number(s) of {\bf internal\/} degrees of freedom of the system(s).

\item the number of internal degrees of freedom of the system, given by 
$H(q_1,\cdots,q_{3r},p_1,\cdots,p_{3r})$~, does generally not correspond to the dimension $k=3r$ of the configuration space.

Given the elements $h_{\mu \nu }$ of the matrix $H$~, the fundamental problem of the theory of matrices consists in solving the eigenvalue equation \cite{v.Neu}, \cite{B-N}:
\[ \sum_\mu h_{\mu \nu }\ s_\nu =\EE_\mu \ s_\mu \;, \qquad 1\le \mu ,\nu \le\infty \;, \]
where:
\Bi
\item the integers $\mu $ and $\nu $ label the internal degrees of freedom,
\item $\EE_\mu $ and $s_\mu $ are respectively the eigenvalues and the corresponding eigenvectors.
\Ei\Ee
\vskip 11pt

\item {\bf Wave aspect\/}

The other attempt of non relativistic quantum mechanics was initiated by L. de Broglie with the idea that, since there exists for the light a corpuscular and a wave aspect related by the energy relation $\EE=h\nu $~, is was natural to suppose that the same duality occurred for the elementary particles to which (periodical) waves had to be associated \cite{Bro1}.

This led him to associate to an elementary particle a wave $\psi $ composed of a superposition of plane waves
\[ \psi =\sum_\mu c(p_\mu )\ e^{i\hbar (\EE_\mu t-p_\mu r)}\]
where $\EE_\mu $ is the energy corresponding to the linear momentum $p_\mu $~.

According to M. Born, the probability of observing an elementary particle with a linear momentum $p$ is given by $|c(p)|^2$ (discrete case).

Following the Hamilton-Jacobi equation of optical geometry, L. de Broglie then proposes the evolution equation:
\[ H(x,y,z,p_x,p_y,p_z)\ \psi =\F\hbar i\ \F{\partial \psi }{\partial t}\]
for the propagation of the wave $\psi $ associated with an elementary particle (in this instance, the electron).

Schrödinger studied extensively the corresponding wave equation:
\[ H(q_1,\cdots,q_{3r},p_1,\cdots,p_{3r})\ \psi (q_1,\cdots,q_{3r})=\lambda \psi (q_1,\cdots,q_{3r})\]
and showed that it was identical to the eigenvalue equation \cite{Vdw}
\[ \sum_\nu h_{\mu \nu }\ s_\nu =\EE_\mu \ s_\mu \]
introduced in b).

However, the above mentioned wave equation is not separable for a system of $r$ elementary particles and, thus, the exact correspondence between the matrix aspect and the wave aspect of the theory is only reached for one isolated elementary particle (or, for an elementary particle (an electron) in the field of a proton: the hydrogen atom studied by E. Schrödinger).  In that case, the rank of the matrix $H(Q_1,Q_2,Q_3,P_1,P_2,P_3)$ to be diagonalized must correspond to the dimension of the basis $\{e^{i\hbar (\EE_\mu t-p_\mu r)}\}_\mu $ in which $\psi $ is developed.
\vskip 11pt

\item {\bf Relativistic aspect}

As it is well known, it is finally P.A.M. Dirac \cite{Dir1} who succeeded in finding the well accepted relativistic wave equation:
\[ \L(\hbar  c\gamma _i\ \F\partial {\partial x^i}+mc^2\R)\ \psi =0\]
which was chosen to be linear in order to have a positive probability density.

This equation has two solutions with positive energy $\EE=+\sqrt{p^2c^2+m^2c^4}$~.  They correspond to the two spin state solutions of an electron with $J _z=\pm\F\hbar 2$~.

The other two solutions refer to the negative energy $\EE=-\sqrt{p^2c^2+m^2c^4}$ and were finally \cite{Dir3} interpreted, in the context of the hole theory, as corresponding to the antiparticle of the electron, the positron \cite{Dir1}, \cite{Dir3}.
\Ee
\vskip 11pt

\section{Second quantization in quantum field theory}

Taking into account the difficulty of interpretation of the hole theory, especially in the case of charged bosons (i.e. the mesons $\pi ^{\pm}$~) \cite{Wei} and the impossibility of developing a relativistic quantum theory with a fixed number of elementary particles, it became necessary to enlarge the frame of relativistic quantum mechanics in order to include a field aspect into the theory \cite{Wig}.

\Bean
\item {\bf Bosonic field}

This was first realized for the radiation field behaving like a sum of independent harmonic oscillators in such a way that each harmonic oscillator in one dimension, characterized by:
\Be
\item the hamiltonian:
\[ H=\half\ (p^2+w^2_0q^2)\]
transformed into
\[ H=\half \ w_0(a^+a+aa^+)=\half\ w_0(a^+_0a_0+a_0a_0^+)\]
if \quad $a=\sqrt{\F1{2w_0}}\ (w_0q+ip)$ \quad and if \quad $a^+=\sqrt{\F1{2w_0}}\ (w_0q-ip)$~,

\item the solutions \quad $a(t)=a_0\ e^{-iw_0t}$ \quad and \quad 
$a^+(t)=a_0^+\ e^{+iw_0t}$ \quad of the equations of motion
\[\dot a(t)=-iw_0a(t) \quad \text{and} \quad \dot a^+(t)=+iw_0a^+(t)\]
coming from \quad $\ddot q+w_0^2q=0$ \quad where \quad $\dot q(t)=\F{dq(t)}{dt}$~,

\item the commutation relations \quad $[a_0,a_0^+]=1$~, \quad $[a_0,a_0]=[a_0^+,a_0^+]=0$~,

\item the eigenvalue equations:
\[ H\psi _\mu =w_\mu \psi _\mu  \quad \text{and}\quad Ha^+_0\psi _\mu =(w_\mu +w_0)a_0^+\psi _\mu \;, \]
\Ee
can generate an infinite set of states of higher energy (starting with a given 
$\psi _\mu $ corresponding to the energy eigenvalue $w_\mu $~) by successive applications 
of the creation operator $a_0^+$~: \quad $a_0^+\psi _\mu =\psi _{\mu +1}$ \quad and a set 
of states of lower energy by successive applications of the annihilation operator 
$a_0$~: \quad $a_0\psi _\mu =\psi _{\mu -1}$ \quad \cite{B-D}, the energy 
$w_\mu $ of the $\mu $-th state $\psi _\mu $ being given by \quad $w_\mu =\L(\mu +\half\R)\ w_0$ \quad where $\half\ w_0$ is the energy of the ground state $\psi _0$~.

The radiation field $u(x,t)$~, solution of the Hamiltonian
\[ H=\half\ \int^L_0\L[\L(\F{\partial u}{\partial t}\R)^2+c^2\ \L(\F{\partial u}{\partial x}\R)^2\R]\ dx\]
can thus be expressed as a sum of Fourier components \cite{Wei}:
\[ u(x,t) = \sum^\infty _{\mu =1}q_\mu (t)\sin \L(\F{w_\mu x}2\R)\]
where the $q$-matrix is given by:
\[ q_\mu (t)=\sqrt{\ds\F\hbar {w_\mu }}\ (a_\mu \exp(-iw_\mu t)+a_\mu ^+\exp (+iw_\mu t))\]
in such a way that the matrix $a_\mu $ or $a_\mu ^+$~, acting on a column vector (with integer components $n_1,n_2,\cdots$~) representing a state with $n_\mu $ quanta in each normal mode $k\equiv \mu $~, lowers or raises the number of quanta $n_\mu $ by one unit.

The Hamiltonian $H$ becomes a sum of oscillator Hamiltonians $H_\mu $ for each cell in momentum space and its diagonal $n$-representation is:
\[(H)_{n'_1,\cdots,n_1}=
\sum_\mu \EE'_\mu 
=\sum_\mu \hbar w_\mu \L(n_\mu +\half\R)\prod_\mu \delta _{n'_\nu n_\nu }\;.\]
It is thus a sum of harmonic oscillators \quad $\EE_\mu =\hbar w_\mu n_\mu $ \quad  plus an infinite zero-point energy \quad $\EE_0=\sum\limits_\mu \ \half\ \hbar w_\mu $~.

This formalism, succinctly recalled for the radiation field, refers to the Bose method which counts the radiation states according to the number $n_\mu $ of quanta in each normal mode.

More specifically, the canonical quantization procedure, applied to the free Klein-Gordon  field, yields a many particle description in terms of numbers of quanta in such a way that an arbitrary state is crudely given by the field:
\[\phi (n_1,\cdots,n_\mu ,\cdots)=\prod_\mu \F1{\sqrt{n_\mu !}}\ (a^+_\mu )^{n_\mu }\phi _\mu (0)\;, \]
where the quanta are indistinguishable since the $a^+_\mu $ commute,\\
and, more exactly, by a symmetric series expansion whose coefficients reflect the symmetry of interchange of quanta in the different normal modes according to the Bose-Einstein statistics \cite{B-D}.
\vskip 11pt

\item {\bf Fermionic field}

On the other hand, the fermionic fields to be quantized were assumed to be relativistic quantum mechanics wave functions in such a way that the informations contained in these do not tell us which particles have which quantum numbers but how many of the indistinguishable particles are in the various quantum modes.  This results from the Pauli exclusion principle preventing the occupation number $n_\mu $ of electrons in any normal mode $\mu $ from taking values other than 0 or 1~.  In this context, the Dirac (electron) field was written according to:
\[ \psi (x)=\sum_\mu a_\mu u_\mu (x)\ 
e^{-iw_\mu t}+\sum_\mu b^+_\mu u_\mu (x)\ e^{+iw_\mu t}\]
where:
\Bi
\item the sum $\sum\limits_\mu $ over the normal modes $\mu $ runs over orthonormal plane-wave solutions of the Dirac equation.
\item $a_\mu $ (resp. $a_\mu ^+$~) are annihilation (resp. creation) operators for positive-energy electrons and $b^+_\mu $ (resp. $b_\mu $~) are annihilation (resp. creation) operators for negative-energy electrons or positrons: they obey anticommutation relations.
\Ei

Correspondingly, the energy operator is:
\[ H=\sum_\mu \hbar w_\mu a_\mu ^+a_\mu +\sum_\mu \hbar |w_\mu |b^+_\mu b_\mu +\EE_0\]
where \quad $\EE_0=-\sum\limits_\mu \hbar |w_\mu |$ \quad is the vacuum energy operator to which corresponds the vacuum state $\psi _0$ containing no positive-energy electrons or positrons.
\Ee\vskip 11pt

\section{Gauge models of the interactions and string theory}

\Bean
\item The {\bf Gauge transformations\/} are based on the observation that there corresponds a conservation law to every continuous symmetry of the Lagrangian in such a way that a transformation on the fields leaving the Lagrangian invariant can be constructed for every conserved quantum number \cite{D-V}.

In quantum electrodynamics, the symmetry operation is a local change of the phase of the electron field, a dephasage resulting from the emission or absorption of a photon.

In the non-abelian electroweak gauge theory of Weinberg-Salam-Glashow, the invariance of the interactions with respect to local transformations of a leptonic equivalent of the isospin generates four fields having null masses, which may become massive by the Higgs mechanism consisting in introducing a new field which doesn't cancel in the vacuum.

So, the vacuum plays an important and complex role in the non-abelian gauge theories where the vacuum state breaks the symmetries obeyed by the equations in order to generate non-vanishing masses while, in quantum electrodynamics, the vacuum state is the zero-particle state.

The quantum chromodynamics is the non-abelian $SU(3)$ gauge theory of colored quarks and gluons which are confined in color singled hadronic bound states: it describes the strong force but does not give a simple qualitative and dynamical understanding of confinement.

Finally, the achievement of the standard model was the elaboration of a unified description of the strong, weak and electromagnetic forces in the context of quantum gauge field theories \cite{G-G-S}.

Unfortunately, at very small distances (Planck length), the quantum fluctuations of the space-time become important breaking down the concept of a continuum space-time: this constitutes the limit of validity of the gauge theories.
\vskip 11pt

\item {\bf String theory}

At the Planck energy (~$\simeq 10^{19}$ Gev~), the standard model is thus falling.  Furthermore, at this energy scale, the gravitational interactions become strong and cannot be neglected.  It was then the challenge of string theory to combine the structure of quantum field theory \cite{Ati} and the standard model with general relativity.

In string theories, point-like particles are replaced by one-dimensional extended strings as fundamental objects in such a way that the basic input parameter is the mass per unit length of the string, its tension \quad $T=\F1{2\pi \alpha '}\equiv\F1{2\pi \ell _s}$ \quad where $\ell _s$ is the characteristic length scale of the string.

In spite of a great activity in superstring theory \cite{DelToWit}, 
\cite{Pol} for several decades, it seems that string theory is not yet a matter field with a stable framework \cite{Wit1}: the underlying conceptual principles are not well understood and, furthermore, there is a lack of contact with experiment \cite{S-S}, \cite{Sch}.
\Ee
\vskip 11pt

\section{The main relevant concepts of quantum (field) theories}

Having quickly reviewed the main concepts of classical, quantum field and string theories, we shall now try to grasp the adequate concepts necessary to build up an algebraic quantum theory whose aim consists in endowing the elementary particles with an internal quantum structure.

So, from the developments of the first and second quantizations, of the gauge and string theories, the following structural concepts may be taken out:
\Bi
\item the dynamics of a set of $r$ particles, described by $r$ material points having {\bbf $k=3r$ external degrees of freedom\/}, is given by a Hamiltonian function of $3r$ coordinates and momenta operators obeying (non-)commutation relations and acting on the particle states. 

{\bf A von Neumann algebra of operators\/} acting on the particle states of a linear {\bf Hilbert space\/} is then introduced \cite{v.Neu}, \cite{Dir2}.

\item the matricial representation of the operators, leading to eigenvalue equations, implies:
\Bean
\item the introduction of {\bf internal dimensions\/} corresponding to the ranks of the matricial representations of the operators.

\item an underlying {\bf concept of bilinearity\/} since the set of $r\times r$ matrices over a ring $R$ forms a $R-R$-bimodule under addition.
\Ee

\item the wave aspect of the first quantization of elementary particles leads to develop the particle mass-wave functions as linear {\bf superpositions of plane waves\/} whose numbers are the above mentioned internal dimensions.

\item the relativistic aspect of the quantum theories, based on {\bf bilinear relativistic invariants\/} of space-time, involves that the solutions of the relativistic equations split into {\bf positive energy solutions of particles\/} and into symmetric {\bf negative energy solutions\/} associated with the corresponding antiparticles \cite{Dir5}.

\item the notion  of field in quantum theories allowed to precise the structure of the 
quantum systems by introducing:
\Bean
\item the radiation field as composed of a set of {\bf harmonic oscillators\/} 
whose (in)finite number corresponds to the quantum internal dimension, also called in 
QFT the {\bf number of normal modes\/}.

\item each normal mode $\mu $ of a harmonic oscillator as composed of 
{\bbf $(n)_\mu $ quanta created from a vacuum state\/}.

\item {\bf creation and annihilation operators\/} respectively raising and lowering the 
numbers of quanta on the harmonic oscillators, allowing to generate an (in)finite set of 
states of higher energy.
\Ee
\Ei
\vskip 11pt

\section{Connecting general relativity to quantum field theories}

The new structure of the proposed algebraic quantum theory will thus be based on the main relevant concepts of quantum field theories, as developed in section 2.5.  It must then be a theory of elementary particles characterized by:
\Bi
\item a quantum nature where the quanta are explicitly described mathematically.
\item a wave aspect.
\item a field and string structure.
\item bilinear invariants of space-time (and of energy-momentum) as those of special relativity.
\Ei

Furthermore, one of the objectives of AQT is the unification of general relativity with quantum field theories at the elementary particle level as developed in \cite{Pie2}.

In this respect, the Einstein field equations:
\[ \lambda g_{\mu \nu }+G_{\mu \nu }=8\pi T_{\mu \nu }\;, \]
where:
\bt[t]{ll}
\textbullet & $\lambda $ is the cosmological constant;\\
\textbullet & $g_{\mu \nu }$ is the metric tensor of space-time;\\
\textbullet  & $G_{\mu \nu }=R_{\mu \nu }-\half\ g_{\mu \nu }R$ with $R_{\mu \nu }$ the Ricci tensor;\\
\textbullet  & $T_{\mu \nu }$ is the stress-energy tensor of matter;
\te\\[11pt]
may receive the following interpretation \cite{Pie7}:
\Bi
\item the vacuum, described by $\lambda g_{\mu \nu }+G_{\mu \nu }=0$~, then corresponds to:
\Bi
\item an expanding space-time structure given by $\lambda g_{\mu \nu }$~;
\item a variation of this internal space-time structure given by 
$G_{\mu \nu }=-\lambda g_{\mu \nu }$ and which must thus be of contracting nature;
\Ei
\item the matter, given by $8\pi T_{\mu \nu }$~, would be generated from the vacuum by the transformation sending \quad $\lambda g_{\mu \nu }+G_{\mu \nu }=0$ \quad into \quad $
\lambda g_{\mu \nu }+G_{\mu \nu }= 8\pi T_{\mu \nu }$~.
\Ei

If we wish to connect general relativity with quantum field theories \cite{P-R}, we have to split the space-time vacuum structure of GR into elementary discrete pieces and consider that these elementary vacua of GR constitute the vacuum fields of QFT from which matter fields can be created.

Thus, the fundamental vacuum fields of AQT, associated with elementary particles, will be of expanding discrete space-time nature.

But, at the macroscopic level of GR, the set of these discrete vacuum fields of elementary particles looks like having a Riemannian continuum space-time structure: this corresponds to a macroscopic limit so that the curvature in the neighbourhood of a point $P$ is equal to the density of matter in this point.

This will constitute the starting point of the developments of AQT whose equations will thus not be derived from a Lagrangian density, as currently done in quantum field theories.  But, the equations of AQT, ``covering'' in some way the equations of QFT, allow to go back to Lagrangian densities.

In this respect, as AQT is not directly connected to Lagrangians having fairly often an ``ad hoc'' character, it will not be a (non abelian) gauge theory.
\vskip 11pt

 \section{Physical tools of AQT}

AQT is a quantum theory of space-time structure of elementary particles.  Its main physical tools will now be succinctly developed and justified.

\Bean
\item Referring to section 2.5, it is assumed that the {\bf fundamental internal structure\/} of an elementary particle is its vacuum structure of space-time.
\vskip 11pt

\item The {\bf relativistic invariants\/} envisaged in AQT as invariants of the space-time structure of elementary particles will not be characterized by a Minkowsky metric as
\begin{alignat*}{5}
dt^2_0 &= c^2\ dt^2-dr^2\;, \qquad && \text{where} \quad &dr^2 &= dx^2+dy^2+dz^2\;, \\
\text{or} \quad
m_0^2c^4&= \EE^2-p^2c^2\;, \qquad && \text{where} \quad &p^2 &= p^2_x+p^2_y+p^2_z\;, \end{alignat*}
but by an euclidian metric, which gives:
\begin{align*}
c^2\ dt^2 &= dt^2_0+dr^2\;, \\
\text{and} \qquad \EE^2 &= m_0^2c^4+p^2c^2\;.\end{align*}
\vskip 11pt

\item Indeed, each one of the three embedded structures, constituting the total structure of an elementary particle as it will be seen, is composed of a {\bf structure of ``space'' type\/}, labeled ``~$S$~'' and of {\bf an orthogonal structure of ``time'' type\/}, labeled ``~$T_0$~'', in such a way that their ``quadratic sum'' \quad $T^2=T_0^2+S^2$ \quad is now an Euclidian invariant of structure.

This is the case since ``~$T_0^2$~'' can be partially or totally transformed into ``~$S^2$~'' and vice versa:
\Bi
\item the case where ``~$T_0^2$~'' is totally transformed into ``~$S^2$~'' corresponds to the annihilation of a fermion pair into a photon (pair).
\item the case where ``~$S^2$~'' is totally transformed into ``~$T_0^2$~'' would correspond to a particle at rest.
\Ei
\vskip 11pt

\item On the other hand, the bilinearity of the relativistic invariants as well as the matricial representation of the operators lead us to consider that the microscopic fundamental structures are twofold: this also results from the solutions of the relativistic wave equations.

In this respect, a new interpretation of the relativistic invariants will consist in considering that every elementary particle is in fact a {\bf bisemiparticle\/} \cite{Pie1}, composed of a left semiparticle, localized in the upper half space, and of a right (symmetric) (co)semiparticle, localized in the lower half space in such a way that:
\Bi
\item the product, right by left, of the right semiparticle by the left semiparticle gives rise to a ``working interaction space'' generating the electric charge and the magnetic moment of the (bisemi)particle by taking into account an off-diagonal metric which, added to the Euclidian metric, leads to a Riemann metric.
\item the right semiparticle, ``dual'' of the left semiparticle, is thus projected on the latter and is unobservable unless the bisemiparticle be split into a pair of ``particle-antiparticle'' when entering into a strong field.
\Ei
\vskip 11pt

\item With this in view, the {\bf space-time structure of the vacuum\/} of a bisemiparticle will be composed of an (internal) time field, corresponding to its ``time'' structure, and of a  space field, corresponding to its ``space'' structure (see c)), in such a way that these fields be of twofold nature and localized in orthogonal spaces.  Referring to the emission and absorption of photons by fermions, it seems judicious to consider that these {\bf time and space fields of the vacua\/} of bisemiparticles, essentially bisemifermions, are of bosonic nature, i.e. composed of a sum of harmonic oscillators characterized by increasing numbers of quanta according to section 2.3 a): this allows to interpret very naturally the quantum jumps and the energy levels of fermions on the basis of their internal structures of vacuum.

Taking into account that an harmonic oscillator can be represented by a pair of by a 
product of two circles having the same radius and rotating in opposite senses 
(see section 4.2 of \cite{Pie3}) and considering the homotopy between a closed 
string and a circle, a vacuum time (or space) field will be given by the (sum of) 
packets of products, right by left, of closed strings in such a way that:
\Bi
\item these packets are characterized by increasing integers $\mu $~, 
$1\le \mu \le q \le \infty $~, referring to the normal modes of a bosonic field.
\item the $\mu $-th packet contains $m_\mu $ products of pairs of closed strings, 
characterized by $\mu $ quanta and localized respectively in the upper and in the 
lower half spaces.
\Ei
\vskip 11pt

\item {\bf The quanta\/}, being irreducible subsets of fields, are assumed to be irreducible algebraic closed subsets \cite{Car} characterized by a Galois extension degree equal to $N$~. Compactified, these quanta constitute ``big points'' of closed strings.
\vskip 11pt

\item As we are concerned with biobjects, we have to consider {\bf biquanta\/} (i.e. products of left quanta by corresponding right quanta) {\bf on bistrings\/} which are products of pairs of right strings localized in the lower half space by the corresponding left strings localized in the upper half space.
\vskip 11pt

\item Remark that the increasing integers $\mu $~, labeling the packets of bistrings and referring to the normal modes of the field, are the {\bf internal dimensions of algebraic nature\/} of the considered system (or field) since they correspond to the numbers of algebraic quanta on the strings.  These integers $\mu $ also refer to the numbers of internal degrees of freedom of a first quantized system according to section 2.2 b) and c).
\vskip 11pt

\item A rotating closed bistring, noted $s_{\mu _R}\times s_{\mu _L}$~, having $\mu $ quanta on $s_{\mu _R}$ and on $s_{\mu _L}$ and belonging to the vacuum space field of a bisemifermion, is interpreted as the vacuum (space) structure of a minimal (bisemi-){\bbf photon at $\mu $ quanta\/}.  Another possibility for a (bisemi-)photon would be $m^{(\mu)} $ closed bistrings at $\mu$ quanta, where $m^{(\mu)} $ denotes the multiplicity, since photons obey the Bose-Einstein statistics.
\vskip 11pt

\item What is especially surprising is the connection of the structure of a field as described in this section with the {\bbf global program of Langlands on $\GL(2)$\/} \cite{Gel}, \cite{Kna}

Indeed, as it will be seen in the next chapter, a field is a (bisemi)sheaf $\widetilde M_R\otimes_D \widetilde M_L$ of $\CC$-valued differentiable bifunctions on the bilinear algebraic semigroup $\GL_2(L_{\o v}\times L_v)$ where $L_v$ (resp. $L_{\o v}$~) denotes (the sum of) the set of real completions corresponding to the \lr ramified algebraic extensions of a global number field of characteristic 0~.  Remark that $\otimes_D$ denotes a ``diagonal'' tensor product characterized by a diagonal metric.

Now, the bisemisheaf $\widetilde M_R\otimes_D \widetilde M_L$ constitutes a representation of the  product, right by left, $W^{ab}_{L_{\o v}}\times 
W^{ab}_{L_{v}}$ of global Weil groups and is in bijection with the cuspidal representation of $\GL_2(\Aa_{L_{\o v}}\times \Aa_{L_{v}})$ where $\Aa_{L_{v}}$ and 
$\Aa_{L_{\o v}}$ are adele semirings over corresponding prime real places: this is the global bilinear correspondence of Langlands on $\GL(2)$~.
\vskip 11pt

\item The space and time fields of the vacuum considered until now are characterized by an Euclidian metric.  If we refer to a Riemann metric, it can then be proved that the off-diagonal components of the metric tensor split into electric and magnetic off-diagonal components to which an electric field, responsible for the {\bf electric charge\/}, and a {\bf magnetic field\/} correspond respectively.

In fact, if $\widetilde M^T_{ST_R}\otimes_D \widetilde M^T_{ST_L}$ denotes the time field of the vacuum and if $\widetilde M^S_{ST_R}\otimes_D \widetilde M^S_{ST_L}$ denotes the corresponding space field, then:
\Bi
\item the off-diagonal tensor product $\widetilde M^S_{ST_R}\otimes_m \widetilde M^S_{ST_L}$ of the vacuum space field is the vacuum magnetic field characterized by a non-orthogonal magnetic metric.
\item the cross tensor products $\widetilde M^T_{ST_R}\otimes_e\widetilde M^S_{ST_L}$ and/or $\widetilde M^S_{ST_R}\otimes_e \widetilde M^T_{ST_L}$ generate(s) the vacuum structure field(s) of the electric charge(s) characterized by a non-orthogonal electric metric.
\Ei
\vskip 11pt

\item The spatial extension of these space and time vacuum fields is of the order of the Planck length.  At this length scale, there are strong fluctuations which generate singularities on the pairs of strings of these fields, or, more exactly, on the pairs of differentiable functions on completions which describe these strings mathematically.

Consequently, versal deformations of degenerate singularities of corank 1 and maximum codimension 3, as well as blowups of these versal deformations are produced in such a way that:
\Bi
\item two embedded fields, labeled by ``~$MG$~'' (for middle-ground) and by ``~$M$~'' (for mass), may cover the time and space fields, labeled by ``~$ST$~'' (for space-time), of the most internal structure of the vacuum of elementary particles according to:
\begin{align*}
\text{time fields:} \quad & \widetilde M^T_{ST_R} \otimes_D \widetilde M^T_{ST_L} 
\subset \widetilde M^T_{MG_R} \otimes_D \widetilde M^T_{MG_L} 
\subset \widetilde M^T_{M_R} \otimes_D \widetilde M^T_{M_L}\;, \\
\text{space fields:} \quad & \widetilde M^S_{ST_R} \otimes_D \widetilde M^S_{ST_L} 
\subset \widetilde M^S_{MG_R} \otimes_D \widetilde M^S_{MG_L} 
\subset \widetilde M^S_{M_R} \otimes_D \widetilde M^S_{M_L} \;.\end{align*}

\item the pairs of closed strings of the time field 
$ \widetilde M^T_{ST_R} \otimes_D \widetilde M^T_{ST_L} $
(resp. space field $ \widetilde M^S_{ST_R} \otimes_D \widetilde M^S_{ST_L} $~)
of the vacuum space-time level, are respectively covered by pairs of open strings of the time (resp. space) fields of the $MG$ and $M$ levels.  Notice that the pairs of strings on the $MG$ and $M$ levels are open strings because, if they have the same number of (bi)quanta as the closed pairs of strings of the $ST$ level that cover, they cannot be closed.
\Ei

The {\bbf mass field ``~$M$~''\/}, 
\[ ( \widetilde M^{TS}_{M_R} \otimes_D \widetilde M^{TS}_{M_L} )
= ( \widetilde M^{T}_{M_R} \otimes_D \widetilde M^{T}_{M_L} ) 
\oplus ( \widetilde M^{S}_{M_R} \otimes_D \widetilde M^{S}_{M_L} )\;,
\]
of a bisemifermion is the (bilinear) field which corresponds to a fermionic field of QFT (see, for example, section 2.3 b)): it {\bf is ``created'' from the vacuum fields\/} $( \widetilde M^{TS}_{MG_R} \otimes_D \widetilde M^{TS}_{MG_L} )
$ and $ ( \widetilde M^{TS}_{ST_R} \otimes_D \widetilde M^{TS}_{ST_L} )$ according to the singularization procedure described above.

The ``~$ST$~'' vacuum field, being presently unobservable, is likely responsible for the {\bf dark energy at the microscopic level\/}.
\vskip 11pt

\item The {\bf bisemifermions\/}, considered in this paper, are the bilinear correspondents of the elementary fermions, that is to say:
\Bi
\item the leptons $e^-$~, $\mu ^-$~, $\tau ^-$ and their neutrinos,
\item the quarks $u^+$~, $d^-$~, $s^-$~, $c^+$~, $b^-$~, $t^+$~.
\Ei

The {\bf bisemihadrons\/}, being the bilinear correspondents of the hadrons composed of baryons and of mesons, are characterized by a central core bistructure of time type to which are tied up three bisemiquarks in the case of bisemibaryons or a pair of (semi)quarks in the case of mesons as it was developed in \cite{Pie4}.

The aim of the physical tools of AQT, reviewed in this section, consists in introducing a plausible internal structure of elementary particles which is summarized in the next proposition.
\Ee
\vskip 11pt

\section{Proposition}

{\em
\Bean
\item Every elementary fermion must be viewed as an elementary bisemifermion:
\Bi
\item composed of  a left semifermion, localized in the upper half space, and of a right semifermion, localized in the symmetric lower half space.
\item centered on an emergence point.
\item to which it can be associated a ``working space'' composed of (tensor) products between right and left internal structures, respectively of the right and of the left semifermions, in such a way that the off-diagonal components of these (tensor) products (after a suitable blow-up morphism) are responsible for the generation of the electric charge and of the magnetic moment.
\Ei
\vskip 11pt

\item An elementary bisemifermion is composed of three central diagonal embedded bistructures whose two internal, labeled ``~$ST$~'' and ``~$MG$~'', are its internal vacuum from which the most external bistructure, which is its mass (``~$M$~'') bistructure, is created.

The vacuum most internal  structure ``~$ST$~'' could correspond to the dark energy at the Planck scale.
\vskip 11pt

\item Each central diagonal bistructure is a (bilinear) field, direct sum of a time (bilinear) field and of a space (bilinear) field.

Each time or space field decomposes into (the sum of) a set of packets of pairs of closed strings in the ``~$ST$~'' case or of open strings in the ``~$MG$~'' and ``~$M$~'' cases.
\vskip 11pt

\item Each packet of pairs of strings:
\Bi
\item is such that the pairs of strings behave like harmonic oscillators.
\item is characterized by an integer $\mu $ corresponding to a normal mode.
\item is such that its strings have a structure composed of $\mu$ quanta which are irreducible algebraic closed subsets of degree $N$~.
\Ei\vskip 11pt

\item Each pair of space field strings, characterized by an integer $\mu $~, is interpreted as the central diagonal bistructure (``~$ST$~, ``~$MG$~'' or ``~$M$~'') of a (bisemi)photon giving then a wave nature of radiation type to the (space) field.
\Ee}

\chapter{Algebraic representations of brane and string fields}

\thispagestyle{empty}

Referring to chapter 2 and, more particularly, to proposition 2.8, the mathematical definition of a time or space (classical) field of the vacuum of a bisemifermion is of central importance.  This will constitute the content of this chapter.
\vskip 11pt


\section{Archimedean symmetric completions}

\Bi
\item Let $K$ be a global number field of characteristic 0 and let $K[x]$ denote a polynomial ring composed of a family of pairs of polynomials $\{P(x),P(-x)\}$~, $x$ being a time or space variable.

The splitting field, generated from $K[x]$~, is the algebraic extension $L^{(c)}$ of $K$~, assumed to be generally closed. This splitting field was shown \cite{Pie5} to be most generally a {\bf symmetric splitting field\/} $ L^{(c)} =L^{(c)} _R\cup L^{(c)} _L$ composed of a right extension semifield $L^{(c)} _R$ and of a left extension semifield $L^{(c)} _L$ in one-to-one correspondence.  The notation $L$ refers to a real splitting field while $L^c$ denotes a complex splitting field.
\vskip 11pt

\item The left and right equivalence classes of Archimedean completions of 
$L^{(c)} _L$ (resp. $L^{(c)} _R$~) are the {\bf left and right places\/} of $L^{(c)} _L$ (resp. $L^{(c)} _R$~) which are such that the real \lr places cover the corresponding complex places: they are noted, in the real case:
\begin{alignat*}{3}
v&= \{v_1,\cdots,v_\mu ,\cdots,v_q\} & \qquad \text{(resp.} \quad
\o v&= \{\o v_1,\cdots,\o v_\mu ,\cdots,\o v_q\}\ )\\
\noalign{\qquad \qquad and, in the complex case:}
\omega &= \{\omega _1,\cdots,\omega _\mu ,\cdots,\omega _q\} & \qquad \text{(resp.} \quad
\o \omega &= \{\o \omega _1,\cdots,\o \omega _\mu ,\cdots,\o \omega _q\}\ ),
\end{alignat*}
\mbox{} \hfill $1\le\mu \le q\le \infty $~.
\vskip 11pt

\item The real {\bf pseudo-ramified completions\/} at the real infinite places $v$ (resp. $\o v$~) are assumed to be generated from {\bbf irreducible one-dimensional $K$-semimodules\/} $L_{v^1_\mu }$ (resp. $L_{\o v^1_\mu }$~) having ranks $[L_{v^1_\mu }:K]=N$ (resp. $[L_{\o v^1_\mu }:K]=N$~) and interpreted as quanta.  The corresponding complex pseudo-ramified completions at the places $\omega $ (resp. $\o\omega $~) are assumed to be generated from irreducible one-dimensional complex $K$-semimodules $L_{\omega ^1_\mu }$ (resp. $L_{\o \omega ^1_\mu }$~) having ranks 
$[L_{\omega ^1_\mu }:L]=m^{(\mu )}\ N$ (resp. $[L_{\o\omega ^1_\mu }:L]=m^{(\mu )}\ N$~), where $m^{(\mu )}=\sup(m_\mu) +1$ is the multiplicity of the $\mu $-th place, in such a way that the complex irreducible completions be covered by the real irreducible completions.

So, the ranks (or degrees) of the real pseudo-ramified completions $L_{v_\mu }$ (resp. $L_{\o v_\mu }$~) will be given by integers modulo $N$ while the ranks of the complex pseudo-ramified completions $L_{\omega _\mu }$ (resp. $L_{\o \omega _\mu }$~) will also be given by integers modulo $N$ according to:
\begin{alignat*}{3}
[L_{v_\mu }:K]
&= *+\mu \centerdot N & \qquad \text{(resp.} \quad 
[L_{\o v_\mu }:K]&= *+\mu \centerdot N \\
&\simeq \mu \ N && \simeq \mu \ N\ )\\
\text{or} \quad 
[L_{\omega _\mu }:K]
&= *+\mu \centerdot m^{(\mu )}\ N & \qquad \text{(resp.} \quad 
[L_{\o \omega _\mu }:K]&= *+\mu \centerdot m^{(\mu )}\ N \\
&\simeq \mu \ m^{(\mu )}\ N && \simeq \mu \ m^{(\mu )}\ N\ )\end{alignat*}
where
\Bi
\item $*$  denotes an integer inferior to $N$~,
\item $\mu $ is called a global residue degree.
\Ei
\vskip 11pt

\item As a place is an 
{\bf equivalence class of completions\/}, 
we have to consider, at each real place $v_\mu $ (resp. $\o v_\mu $~), 
a set of $m^{(\mu )}$ real completions $L_{v_{\mu ,m_\mu }}$ 
(resp. $L_{\o v_{\mu ,m_\mu }}$~), $m_\mu \in\NN$~, $m^{(\mu )}=\sup(m_\mu )+1$~, 
equivalent to $L_{v_\mu }$ (resp. $L_{\o v_\mu }$~), with $m_\mu =0$~, 
and characterized by the same ranks as $L_{v_\mu }$ (resp. $L_{\o v_\mu }$~).

On the other hand, as the complex completions were assumed to be covered by the real 
completions, the multiplicity $m^{(\mu )}$ of the complex completions will be 
equal to 0~, $\forall\ \mu $~, $1\le \mu \le q\le \infty $~.
\vskip 11pt

\item Let $L_{v_+}= \bigoplus\limits_\mu L_{v_\mu } \bigoplus\limits_{m_\mu} L_{v_{\mu ,m_\mu }} $ 
(resp. $L_{\o v_+}= \bigoplus\limits_\mu L_{\o v_\mu } \bigoplus\limits_{m_\mu} L_{\o v_{\mu ,m_\mu }} $~) denote the {\bf sum of the real completions\/} at all places of $L_L$ (resp. $L_R$~), and let $L_{\omega _+}= \bigoplus\limits_\mu L_{\omega _\mu} $
(resp. $L_{\o \omega _+}= \bigoplus\limits_\mu L_{\o \omega _\mu} $~) be the corresponding sum of complex completions of $L^c_L$ (resp. $L_R^c$~).

In this context, the {\bf pseudo-ramified adele semiring\/} $\Aa_{L_v}$ (resp. $\Aa_{L_{\o v}}$~) will be introduced in the real case by:
\begin{alignat*}{3}
\Aa_{L_v} &= \prod\limits_{\mu _p} L_{v_{\mu _p}}
\prod\limits_{m_{\mu _p}} L_{v_{\mu _p,m_{\mu _p}}}\qquad
&\text{(resp.} \quad 
\Aa_{L_{\o v}} =& \prod\limits_{\mu _p} L_{\o v_{\mu _p}}
\prod\limits_{m_{\mu _p}} L_{\o v_{\mu _p,m_{\mu _p}}}\ )\\
\noalign{\qquad \qquad and, in the complex case, by:}
 \Aa_{L_\omega }&=\prod\limits_{\mu _p} L_{\omega _{\mu _p}}\qquad
\qquad &\text{(resp.} \quad 
\Aa_{L_{\o \omega }}=& \prod\limits_{\mu _p} L_{\o \omega _{\mu _p}}\ )\end{alignat*}
where the product $\prod\limits_{\mu _p}$ runs over the Archimedean prime completions \cite{J-L}.
\Ei
\vskip 11pt

\section{Algebraic bilinear semigroups over real completions}

\Bi
\item Let $T_2(L_v)$ (resp. $T_2^t(L_{\o v})$~) denote the group of upper (resp. lower) triangular matrices of order 2 over the set $L_v=\{ L_{v_1}, \cdots,
L_{v_\mu }, \cdots, L_{v_{\mu ,m_\mu }}, \cdots, L_{v_{q,m_q}}\}$ (resp.
$L_{\o v}=\{ L_{\o v_1}, \cdots,
L_{\o v_\mu }, \cdots, L_{\o v_{\mu ,m_\mu }}, \cdots,\linebreak L_{\o v_{q,m_q}}\}$~) of real completions.
\vskip 11pt

\item Then, an algebraic bilinear general semigroup {\bbf $GL_2(L_{\o v}\times L_v) =
T_2^t(L_{\o v})\times T_2(L_v)$} can be introduced in such a way that:
\Bean
\item the product $(L_{\o v}\times L_v)$ over the two sets 
$L_{\o v}$ and $L_v$ of completions must be taken over the set 
$\{L_{\o v_{\mu ,m_\mu }}\times L_{v_{\mu ,m_\mu }}\}_{v_{\mu ,m_\mu }}$ 
of products of corresponding 
pairs of completions.

\item $\GL_2(L_{\o v}\times L_v)$ has the Gauss bilinear decomposition:
\[  \GL_2(L_{\o v}\times L_v)
= [D_2(L_{\o v}) \times D_2(L_v)][UT_2(L_v)\times UT_2^t(L_{\o v})]\]
where:
\Bi
\item $D_2(\centerdot)$ is a subgroup of diagonal matrices.
\item $UT_2(\centerdot)$ (resp. $UT^t_2(\centerdot)$~) is the subgroup of upper (resp. lower) unitriangular matrices.
\Ei
\vskip 11pt

\item $\GL_2(L_{\o v}\times L_v)$ has for modular representation space 
$\Repsp (\GL_2(L_{\o v}\times L_v))$ the tensor product 
$M_R(L_{\o v}) \otimes M_L(L_v)$ of a right $T_2^t(L_{\o v})$-semimodule  
$M_R(L_{\o v}) $ by a left $T_2(L_v)$-semimodule $M_L(L_v)$~, also noted $M_R\otimes M_L$~.
\vskip 11pt

\item $\GL_2(L_{\o v}\times L_v)$ covers its linear equivalent $\GL_2(L_{\o v-v})$ \cite{Bor}, where $L_{\o v-v}\simeq L_{\o v}\cup L_v$~, having the linear Gauss decomposition:
\[  \GL_2(L_{\o v-v})
= D_2(L_{\o v-v}) \times[UT_2(L_{\o v-v})\times UT_2^t(L_{\o v-v})]\]
if we take into account the maps:
\Bi
\item $UT_2(L_{\o v-v})\to UT_2(L_v)$~.
\item $ UT_2^t(L_{\o v-v})\to UT_2^t(L_{\o v})$~.
\item $D_2(L_{\o v-v})\to D_2 (L_{\o v}\times L_v)$~.
\Ei
\vskip 11pt

\item its $\mu $-th conjugacy class representative with respect to the product, 
right by left, $L_{\o v^1_\mu }\times L_{v^1_\mu }$ of irreducible real completions 
of rank $N$ has for representation the 
$\GL_2(L_{\o v_{\mu,m_\mu } } \times L_{v_{\mu,m_\mu } } )$-subbisemimodule 
$M_{\o v_{\mu ,m_\mu }} \otimes M_{v_{\mu ,m_\mu }} $ where 
$M_{v_{\mu ,m_\mu }} $ (resp. $M_{\o v_{\mu ,m_\mu }} $~) constitutes the 
one-dimensional modular representation  of the $(\mu ,m_\mu )$-th conjugacy class 
representative of $T_2(L_v)$ (resp. $T_2^t(L_{\o v})$~).

In the context of QFT, $M_{\o v_{\mu ,m_\mu }} $ and $M_{v_{\mu ,m_\mu }} $ are strings at $\mu $ quanta.
\Ee\vskip 11pt

\item An algebraic bilinear semigroup {\bbf $\GL_2(L_{\o v_+ } \times L_{v_+ } )$} over the product of the sums
\[
L_{\o v_+} = \bigoplus_\mu L_{\o v_\mu } \bigoplus_{m_\mu } L_{\o v_{\mu,m_\mu } } 
\qquad
\text{and} \qquad 
L_{v_+} = \bigoplus_\mu L_{v_\mu } \bigoplus_{m_\mu } L_{v_{\mu,m_\mu } } 
\]
of real completions, has for modular representation  space
$\Repsp (\GL_2(L_{\o v_+ } \times L_{v_+ } ))$ the tensor product 
$M_R(L_{\o v_+ } )\otimes M_L(L_{v_+})$~, also written 
$M_R^+\otimes M_L^+$~, of a right $T_2^t(L_{\o v_+})$-semimodule 
$M_R^+$ by a left $T_2(L_{v_+})$-semimodule $M_L^+$~.

$M_R^+\otimes M_L^+$~, which is a $\GL_2(L_{\o v_+ } \times L_{v_+ } )$-bisemimodule, decomposes according to:
\[ M_R^+\otimes M_L^+= \bigoplus^q_{\mu =1} \bigoplus_{m_\mu } (
M_{\o v_{\mu ,m_\mu }}\otimes M_{v_{\mu ,m_\mu }} )\]
$M^+_R$ (resp. $M_L^+$~) has a rank $n_R$ (resp. $n_L$~) given by:
\[ n_R\equiv n_L=\sum_\mu \sum_{m_\mu }(\mu \times N)\]
if it is referred to section 3.1.
\vskip 11pt

\item Finally, an algebraic bilinear semigroup {\bbf $\GL_2(\Aa_{L_{\o v}}\times \Aa_{L_v})$} over the product of adele semirings $\Aa_{L_{\o v}}$ and
$\Aa_{L_{v}}$ has for representation space 
$\Repsp(GL_2(\Aa_{L_{\o v}}\times \Aa_{L_v}))$ the tensor product 
$M_R(\Aa_{L_{\o v}} )\otimes M_L(\Aa_{L_{v}})$ of a right $T_2^t(\Aa_{L_{\o v}})$-semimodule $M_R(\Aa_{L_{\o v}})$ by a left 
$T_2(\Aa_{L_{v}})$-semimodule $M_L(\Aa_{L_{v}})$ in such a way that
$\GL_2(\Aa_{L_{\o v}}\times \Aa_{L_v})$ may have $M_R^+\otimes M_L^+$ as a modular representation space if the composition of (bi)homomorphisms:
\[
\begin{CD}
\GL_2(\Aa_{L_{\o v}}\times \Aa_{L_v}) @>>> 
M_R(\Aa_{L_{\o v}} )\otimes M_L(\Aa_{L_{v}})\\
@VVV @VVV\\
\GL_2(L_{\o v_+ } \times L_{v_+ } ) @>>> M_R^+\otimes M^+_L\end{CD}\]
is taken into account.
\Ei
\vskip 11pt

\section{Algebraic bilinear semigroups over complex completions}

\Bi
\item Let
\[
L_\omega = \{L_{\omega _1},\cdots,L_{\omega _\mu },\cdots,L_{\omega _q}\}\qquad
\text{(resp.}\quad
L_{\o\omega} = \{L_{\o \omega _1},\cdots,L_{\o \omega _\mu },\cdots,L_{\o \omega _q}\}\ )\]
be the set of complex completions covered by the set of real completions $L_v$ (resp. $L_{\o v}$~).

Then, similarly as in section 3.2, an algebraic bilinear semigroup 
{\bbf $GL_2(L_{\o \omega }\times L_\omega ) \equiv
T_2^t(L_{\o \omega })\times T_2(L_\omega )$} over products of corresponding pairs of complex completions can be introduced in such a way that:
\Bean
\item $GL_2(L_{\o \omega }\times L_\omega ) $ has a Gauss bilinear decomposition.
\item $GL_2(L_{\o \omega }\times L_\omega ) $ has for modular representation space
 $\Repsp(GL_2(L_{\o \omega }\times L_\omega ) )$ the tensor product $M_R(L_{\o\omega })\otimes M_L(L_\omega )$ of a right $T_2^t(L_{\o\omega })$-semimodule 
$M_R(L_{\o\omega })$ by a corresponding symmetric left $T_2(L_\omega )$-semimodule $M_L(L_\omega )$~.

\item $GL_2(L_{\o \omega }\times L_\omega ) $  covers its linear equivalent
$GL_2(L_{\o \omega -\omega}) $ where $L_{\o\omega -\omega }\simeq L_{\o\omega }\cup L_\omega $~.

\item Its $\mu $-th conjugacy class (representative) with respect to the product, right by left, $L_{\o \omega ^1_\mu }\times L_{\omega ^1_\mu }$ of irreducible complex completions of rank $N$ has for representation the 
$GL_2(L_{\o \omega_\mu  }\times L_{\omega_\mu } ) $-subbisemimodule
$M_{\o \omega_\mu  }\otimes M_{\omega_\mu } $ where $M_{\o \omega_\mu  }$ (resp.
$M_{\omega_\mu  }$~) is the one-dimensional complex representation of the $\mu $-th conjugacy class of $T_2(L_\omega )$ (resp. $T_2^t(L_{\o\omega }$~).

\item In the context of string theory, $M_{\omega _\mu }$ and $M_{\o \omega _\mu }$ would be branes at $\mu \times m^{(\mu )}$ quanta according to section 3.1.
\Ee
\vskip 11pt

\item An algebraic bilinear semigroup {\bbf $\GL_2(L_{\o \omega _+ } \times L_{\omega _+ } )$} over the product of the sums
\[
L_{\o \omega _+} = \bigoplus_\mu L_{\o \omega _\mu } \qquad 
\text{and} \qquad 
L_{\omega _+} = \bigoplus_\mu L_{\omega _\mu } \]
of complex completions, has for modular representation  space
$\Repsp (\GL_2(L_{\o \omega _+ } \times L_{\omega _+ } ))$ the tensor product 
$M^+_R(L_{\o \omega _+ } )\otimes M^+_L(L_{\omega _+})$ of a right 
$T_2^t(L_{\o \omega_+ })$-semimodule 
$M_R^+(L_{\o\omega _+})$ by a left $T_2(L_{\omega _+})$-semimodule 
$M_L^+(L_{\omega _+})$~.

$M_R^+(L_{\o\omega _+})\otimes M_L^+(L_{\omega _+})$ decomposes into:
\[ M_R^+(L_{\o\omega _+})\otimes M_L^+(L_{\omega _+})
=\bigoplus^q_{\mu =1} (
M_{\o \omega _{\mu  }}\otimes M_{\omega _{\mu }} )\]
where $M^+_{\omega _\mu }$ and $M^+_{\o\omega _\mu }$ have a rank 
\[ n_{\omega _\mu ^+} =\mu \times m^{(\mu )}\times N\;.\]
\vskip 11pt

\item An algebraic bilinear semigroup {\bbf $\GL_2(\Aa_{L_{\o \omega }}\times \Aa_{L_\omega })$} over the product of adele semirings $\Aa_{L_{\o \omega }}$ and
$\Aa_{L_{\omega }}$ has for representation space 
$\Repsp(GL_2(\Aa_{L_{\o \omega }}\times \Aa_{L_\omega }))$ the tensor product 
$M_R(\Aa_{L_{\o \omega }} )\otimes M_L(\Aa_{L_{\omega }})$ of a right $T_2^t(\Aa_{L_{\o \omega }})$-semimodule $M_R(\Aa_{L_{\o \omega }})$ by a left symmetric 
$T_2(\Aa_{L_{\omega }})$-semimodule $M_L (\Aa_{L_{\omega }})$ in such a way that
$M_R(\Aa_{L_{\o \omega }} )\otimes M_L(\Aa_{L_{\omega }})$ may have the bisemimodule $M_R^+(L_{\o\omega _+})\otimes M_L^+(L_{\omega _+})$ as  modular representation space if the composition of (bi)homomorphisms:
\[
\begin{CD}
\GL_2(\Aa_{L_{\o \omega }}\times \Aa_{L_\omega }) @>>> 
M_R(\Aa_{L_{\o \omega }} )\otimes M_L(\Aa_{L_{\omega }})\\
@VVV @VVV\\
\GL_2(L_{\o \omega _+ } \times L_{\omega _+ } ) @>>> M_R^+
({L_{\o\omega _+}})\otimes M^+_L(({L_{\omega_+ }})\end{CD}\]
is considered.
\Ei
\vskip 11pt

\section{Toroidal compactifications}

 {\bf A toroidal compactification\/} of the real and complex completions must then be envisaged in such a way that the real completions are transformed into one-dimensional (semi)tori or (semi)circles and the complex completions are transformed into two-dimensional (semi)tori.  This toroidal compactification was introduced in chapter 1 of \cite{Pie4} and corresponds to a projective emergent toroidal isomorphism of completions:
\begin{alignat*}{3}
\gamma ^{(1)}_{\mu _L} : \quad L_{v_\mu } & \longrightarrow L^T_{v_\mu }
 \qquad \qquad 
\text{(resp.} \quad 
&\gamma ^{(1)}_{\mu _R} : \quad L_{\o v_\mu } & \longrightarrow L^T_{\o v_\mu }\ )\\
\gamma ^{(2)}_{\mu _L} : \quad L_{\omega _\mu } & \longrightarrow L^T_{\omega _\mu }
 \qquad \qquad 
\text{(resp.} \quad 
&\gamma ^{(2)}_{\mu _R} : \quad L_{\o \omega _\mu } & \longrightarrow L^T_{\o \omega _\mu }\ ).\end{alignat*}
These toroidal compactifications of completions then involve the homomorphisms of algebraic bilinear semigroups:
\begin{alignat*}{5}
H_{\o v-v} 
&: \quad \GL_2(L_{\o v}\times L_v)
&\longrightarrow & \GL_2 (L^T_{\o v}\times L^T_v)
&\simeq & O_2 (L^T_{\o v}\times L^T_v)\; ,\\
H_{\o v_+-v_+} 
&: \quad \GL_2(L_{\o v_+}\times L_{v_+})
&\longrightarrow & \GL_2 (L^T_{\o v_+}\times L^T_{v_+})
&\simeq & O_2 (L^T_{\o v_+}\times L^T_{v_+})\; ,\\
H_{\Aa_{L_{\o v-v}} }
&: \quad \GL_2(\Aa_{L_{\o v}}\times \Aa_{L_v})
&\longrightarrow & \GL_2 (\Aa_{L^T_{\o v}}\times \Aa_{L^T_v})
&\simeq & O_2 (\Aa_{L^T_{\o v}}\times \Aa_{L^T_v})\; ,\\
H_{\o \omega -\omega } 
&: \quad \GL_2(L_{\o \omega }\times L_\omega )
&\longrightarrow & \GL_2 (L^T_{\o \omega }\times L^T_\omega )
&\simeq & U_2 (L^T_{\o \omega }\times L^T_\omega )\; ,\\
H_{\o \omega_+ -\omega_+ } 
&: \quad \GL_2(L_{\o \omega _+}\times L_{\omega_+} )
&\longrightarrow & \GL_2 (L^T_{\o \omega_+ }\times L^T_{\omega _+})
&\simeq & U_2 (L^T_{\o \omega_+ }\times L^T_{\omega_+} )\; ,\\
H_{\Aa_{L_{\o \omega -\omega }} }
&: \quad \GL_2(\Aa_{L_{\o \omega }}\times \Aa_{L_\omega} )
&\longrightarrow & \GL_2 (\Aa_{L^T_{\o \omega }}\times \Aa_{L^T_\omega} )
&\simeq & U_2 (\Aa_{L^T_{\o \omega }}\times \Aa_{L^T_\omega} )\; ,\end{alignat*}

where $O_2(\centerdot_R\times \centerdot_L)$ is the bilinear orthogonal (semi)group which may be introduced by setting:
\[O_2(\centerdot_R\times \centerdot_L) = O^T_2(\centerdot_R)^{-1}\times O_2(\centerdot_L);.\]
\vskip 11pt

\section{Inclusions of the ``real'' bilinear algebraic semigroups into their ``complex'' equivalents}

\Bi
\item Let $M_R(L_{\o\omega })\otimes M_L(L_{\omega }))$ be the representation space 
$\Repsp (\GL_2(L_{\o\omega }\times L_\omega ))$ of the bilinear algebraic semigroup
$\GL_2(L_{\o\omega }\times L_\omega )$ over products of pairs of complex completions.
\vskip 11pt

\item And let $M_R(L_{\o v})\otimes M_L(L_{v})$ be the corresponding representation space 
$\Repsp (\GL_2(L_{\o v}\times L_v))$ of the bilinear algebraic semigroup
$\GL_2(L_{\o v}\times L_v)$ over products of pairs of real completions.
\vskip 11pt

\item The inclusion $M_R(L_{\o v})\otimes M_L(L_{v})\subseteq
M_R(L_{\o\omega })\otimes M_L(L_{\omega })$ of the real
$\GL_2(L_{\o v}\times L_v)$-bisemimodule $M_R(L_{\o v})\otimes M_L(L_v)$ into the complex $\GL_2(L_{\o \omega }\times L_\omega )$-bisemimodule $M_R(L_{\o \omega })\otimes M_L(L_\omega )$ implies that:
\Bi\item each $\mu $-th complex conjugagy class representative $M_{\o\omega _\mu }\otimes M_{\omega _\mu }$~, isomorphic to its toroidal equivalent  
$M^T_{\o\omega _\mu }\otimes M^T_{\omega _\mu }$~, is covered by the set of $m^{(\mu )}=\sup(m_\mu )+1$ real conjugacy class representatives $\{
M_{\o v_{\mu ,m_\mu }} \otimes M_{v_{\mu ,m_\mu }} \}_{m_\mu }$~, isomorphic to their toroidal equivalents 
$\{
M^T_{\o v_{\mu ,m_\mu }} \otimes M^T_{v_{\mu ,m_\mu }} \}_{m_\mu }$~,
$
M_{\o \omega _{\mu  }} \otimes M_{\omega _{\mu  }} \in M_R(L_{\o\omega })\otimes M_L(L_\omega )$
and $
M_{\o v_{\mu,m_\mu   }} \otimes M_{v_{\mu,m_\mu   }}   \in M_R(L_{\o v})\otimes M_L(L_v)$~.

\item each $\mu $-th complex conjugacy class representative 
$M_{\o\omega _\mu }\otimes M_{\omega _\mu }$ is composed of $\mu $ equivalent 
conjugacy class subrepresentatives $M_{\o\omega ^{\mu '}_\mu}  \otimes 
M_{\omega ^{\mu '}_\mu }$~, $1\le \mu '\le \mu $~, 
of which $M_{\omega ^{\mu '}_\mu}  $ (resp. $M_{\o\omega ^{\mu '}_\mu}$~) has a rank $N\times m^{(\mu )}$~, and each $(\mu ,m_\mu )$-th real conjugacy class representative
$M_{\o v_{\mu ,m_\mu }}\otimes  M_{v_{\mu ,m_\mu }}$ is composed of $\mu $ 
equivalent real conjugacy class subrepresentatives 
$M_{\o v^{\mu ^{''}_\mu}}  \otimes M_{v^{\mu^{''}_\mu }}$~, 
$1\le \mu ^{''}\le \mu $~, 
of which $M_{v^{\mu ^{''}_\mu}}  $ (resp. $M_{\o v^{\mu ^{''}_\mu}}$~) has a rank 
$N$ and is a quantum, in such a way that every 
$M_{\o\omega ^{\mu '}_\mu}  \otimes M_{\omega ^{\mu '}_\mu }$ is covered by $m^{(\mu )}$ 
{\bf biquanta\/} $M_{\o v^{\mu^{''}_\mu} } \otimes M_{v^{\mu ^{''}_\mu}}$ 
according to section 3.3.
\Ei\Ei
\vskip 11pt

\section{(Bisemi)Sheaves over algebraic bilinear semigroups}

\Bi
\item Let $\phi  _L(M_{\omega _\mu })$ (resp. $\phi _R(M_{\o\omega _\mu })$~) denote a $\CC$-valued differentiable function over the $\mu $-th complex conjugacy class representative $M_{\omega _\mu }$ (resp. $M_{\o\omega _\mu }$~) of $T_2(L_\omega )$ (resp. $T^t_2(L_{\o\omega })$~) $\subset \GL_2(L_{\o\omega }\times L_\omega )$~.

The tensor product $\phi  _R(M_{\o\omega _\mu })\otimes \phi  _L(M_{\omega _\mu })$ called a $\CC$-valued differentiable bifunction:
\Bi
\item verifies $(\phi _R\otimes\phi _L)(M_{\o\omega _\mu }\otimes 
M_{\omega _\mu })
=  \phi  _R(M_{\o\omega _\mu })\otimes \phi  _L(M_{\omega _\mu })$~.

\item is defined over the $\mu $-th conjugacy class representative
$ M_{\o\omega _\mu }  \otimes M_{\omega _\mu }  $ of $\GL_2(L_{\o\omega }  \times L_\omega )$~.
\Ei
\vskip 11pt

\item   Let $\phi  _L(M_{v_\mu })$ (resp. $\phi _R(M_{\o v_\mu })$~) be a complex-valued differentiable function over the $\mu $-th real conjugacy class representative
$M_{v_\mu }$ (resp. $M_{\o v_\mu }$~) of $T_2(L_v)$ (resp. $T^t_2(L_{\o v})$~) $\subset \GL_2(L_{\o v}\times L_v)$ and let 
$\phi _R(M_{\o v_\mu })\otimes \phi _L(M_{v_\mu })$ denote the corresponding bifunction over the conjugacy class representative 
$M_{\o v_\mu }\otimes  M_{v_\mu }$ of $\GL_2(L_{\o v}\times L_v)$~.
\vskip 11pt

\item The set $\{\phi _L(M_{\omega_\mu })\}^q_{\mu =1}$ (resp.
$\{\phi _R(M_{\o\omega_\mu })\}^q_{\mu =1}$~) of differentiable functions, localized in the upper (resp. lower) half space and defined over the $T_2(L_\omega )$ (resp. $T^t_2(L_{\o \omega })$~)-semi\-mo\-du\-le $M_L(L_\omega )$ (resp. $M_R(L_{\o\omega })$~), constitutes the set $\Gamma (\phi _L(M_L(L_\omega )))$ (resp.
$\Gamma (\phi _R(M_R(L_{\o\omega} )))$~) of sections of a {\bf semisheaf of rings\/} (or a sheaf of semirings!) {\bbf $\phi _L(M_L(L_\omega ))$ (resp.
$  \phi _R(M_R(L_{\o\omega} )) $~)\/}, as introduced in \cite{Pie4}.

And, the set $\{ \phi _R(M_{\o\omega_\mu }) \otimes \phi _L(M_{\omega_\mu }) \}^q_{\mu =1}$ of differentiable bifunctions over the $\GL_2(L_{\o\omega }\times L_\omega )$-bisemi\-mo\-dule $M_R(L_{\o\omega })\otimes M_L(L_\omega )$ constitutes the set
$\Gamma (\phi _R(M_R(L_{\o\omega} ))\otimes \phi _L(M_L(L_\omega )))$  of bisections of a {\bbf bisemisheaf of rings  $ \phi _R(M_R(L_{\o\omega} ))\otimes \phi _L(M_L(L_\omega ))$}~.
\vskip 11pt

\item Similarly, the set $\{\phi _L(M_{v_{\mu,m_\mu } })\}_{\mu,m_\mu }$ (resp.
$\{\phi _R(M_{\o v_{\mu,m_\mu } })\}_{\mu,m_\mu }$~) of $\CC$-valued differentiable functions, localized in the upper (resp. lower) half space and defined over the $T_2(L_v)$ (resp. $T^t_2(L_{\o v})$~)-semimodule $M_L(L_v)$ (resp. $M_R(L_{\o v})$~), constitutes the set $\Gamma (\phi _L(M_L(L_v)))$ (resp. $\Gamma (\phi _R(M_R(L_{\o v})))$~) of sections of a {\bbf semisheaf of rings $ \phi _L(M_L(L_v))$ 
(resp. $ \phi _R(M_R(L_{\o v}))$~)}.

And, the set $\{ \phi _R(M_{\o v_{\mu,m_\mu } }) 
\otimes \phi _L(M_{v_{\mu ,m_\mu }}) \}_{\mu ,m_\mu }$ of differentiable bifunctions 
over the $\GL_2(L_{\o v}\times L_v)$-bisemimodule $M_R(L_{\o v})\otimes M_L(L_v)$ 
constitutes the set
$\Gamma (\phi _R(M_R(L_{\o v} ))\otimes \phi _L(M_L(L_ v)))$  of bisections of the 
{\bbf bisemisheaf of rings  $ \phi _R(M_R(L_{\o v} ))\otimes \phi _L(M_L(L_v))$}~.
\Ei
\vskip 11pt

\section{Proposition}

{\em The real bisemisheaf of rings $ \phi _R(M_R(L_{\o v }))
\otimes \phi _L(M_L(L_{v }))$ is included into the complex corresponding bisemisheaf of rings $ \phi _R(M_R(L_{\o \omega  }))
\otimes \phi _L(M_L(L_{\omega  }))$~.}
\vskip 11pt

\bpr Indeed, every complex conjugacy class representative $M_{\o\omega _\mu }\otimes M_{\omega _\mu }$ over which is defined a bisection 
$ \phi _R(M_{\o \omega _\mu  })
\otimes \phi _L(M_{\omega _\mu  })$ of the bisemisheaf 
$ \phi _R(M_R(L_{\o \omega }))
\otimes \phi _L(M_L(L_{\omega  }))$ is covered by the set
$\{M_{\o v_{\mu ,m_\mu }}\otimes M_{v_{\mu ,m_\mu }}\}_{m_\mu }$ of real conjugacy 
class representatives over which are defined the set 
$\{\phi _R(M_{\o v_{\mu ,m_\mu }}\otimes\linebreak \phi _L(M_{v_{\mu ,m_\mu }})\}_{m_\mu }$ of $m^{(\mu )}$ bisections of the bisemisheaf $\phi _R(M_R(L_{\o v}))\otimes \phi _L(M_L(L_v))$~.\epr
\vskip 11pt

\section{Bisemimodules associated with bisemisheaves}

\Bi
\item If we take the direct sum 
$\bigoplus^q_{\mu =1}(\phi _R(M_{\o\omega _\mu }) \otimes \phi _L(M_{\omega _\mu }))$ of all bisections of the complex bisemisheaf $\phi _R(M_R(L_{\o \omega })) \otimes \phi _L(M_L(L_\omega ))$~, we get a
$\GL_2(L_{\o\omega _+}\times L_{\omega _+})$-bisemimodule 
$\phi _R(M^+_R(L_{\o \omega_+ })) \otimes \phi _L(M^+_L(L_{\omega _+}))$~.
\vskip 11pt

\item Similarly, the direct sum
$\bigoplus^q_{\mu =1}\bigoplus_{m_\mu } (\phi _R(M_{\o v_{\mu,m_\mu } }) \otimes 
\phi _L(M_{v_{\mu ,m_\mu} }))$ of all bisections of the real bisemisheaf $\phi _R(M_R(L_{\o v})) \otimes \phi _L(M_L(L_v))$ generates a
$\GL_2(L_{\o v_+}\times L_{v_+})$-bisemimodule 
$\phi _R(M^+_R(L_{\o v_+ })) \otimes \phi _L(M^+_L(L_{v_+}))$~.
\vskip 11pt

\item On the other hand, the direct product
$\prod_{\mu _p} (\phi _R(M_{\o \omega _{\mu_p  }}) \otimes \phi _L(M_{\omega _{\mu _p} }))$ of all ``primary'' bisections of the complex bisemisheaf $\phi _R(M_R(L_{\o \omega })) \otimes \phi _L(M_L(L_\omega ))$ gives rise to  a
$\GL_2(\Aa_{L_{\o \omega }}\times \Aa_{L_{\omega }})$-bisemimodule 
$\phi _R(M_R(\Aa_{L_{\o \omega }})) \otimes \phi _L(M_L(\Aa_{L_{ \omega }}))$~.
\vskip 11pt

\item And the direct product
$\prod_{\mu _p,m_{\mu _p}} (\phi _R(M_{\o v_{\mu_p,m_{\mu _p}  }}) \otimes \phi _L(M_{v_{\mu _p,m_{\mu _p}} }))$ of all primary bisections of the real bisemisheaf $\phi _R(M_R(L_{\o v})) \otimes \phi _L(M_L(L_ v))$ generates a
$\GL_2(\Aa_{L_{\o v}}\times \Aa_{L_{v}})$-bisemimodule 
$\phi _R(M_R(\Aa_{L_{\o v}})) \otimes \phi _L(M_L(\Aa_{L_{ v}}))$~.
\vskip 11pt

\item In the same way, over the toroidally compactified completions, the
\Bi
\item $\GL_2(L^T_{\o\omega _+}\times L^T_{\omega _+})$-bisemimodule
$\phi _R(M^+_R(L^T_{\o\omega _+})) \otimes \phi _L(M^+_L(L^T_{\omega _+})) $~,

\item $\GL_2(L^T_{\o v_+}\times L^T_{v_+})$-bisemimodule
$\phi _R(M^+_R(L^T_{\o v_+})) \otimes \phi _L(M^+_L(L^T_{v_+})) $~,

\item $\GL_2(\Aa_{L^T_{\o\omega }}\times \Aa_{L^T_{\omega }})$-bisemimodule
$\phi _R(M_R(\Aa_{L^T_{\o\omega}})) \otimes \phi _L(M_L(\Aa_{L^T_{\omega }})) $~,

\item $\GL_2(\Aa_{L^T_{\o v}}\times \Aa_{L^T_{v}})$-bisemimodule
$\phi _R(M_R(\Aa_{L^T_{\o v}})) \otimes \phi _L(M_L(\Aa_{L^T_{v}})) $~,
\Ei
corresponding respectively to the above-defined bisemimodules, can be introduced.
\Ei\vskip 11pt

\section{Proposition}

{\em
\Be
\item The complex bisemisheaf of rings $ \phi _R(M_R(L_{\o\omega }))\otimes
\phi _L(M_L(L_{\omega })$ is a physical ``{\/\bf brane field\/}'' having representations in the:
\Be
\item bisemimodule 
$\phi _R(M^+_R(L_{\o\omega_+ }))\otimes \phi _L(M^+_L(L_{\omega_+ }))$ 
isomorphic to its toroidal equivalent\linebreak
$\phi _R(M^+_R(L^T_{\o\omega_+ }))\otimes \phi _L(M^+_L(L^T_{\omega_+ }))$~;
\item bisemimodule 
$\phi _R(M_R(\Aa_{L_{\o\omega} }))\otimes \phi _L(M_L(\Aa_{L_{\omega} }))$ 
isomorphic to its toroidal equivalent\linebreak
$\phi _R(M_R(\Aa_{L^T_{\o\omega} }))\otimes \phi _L(M_L(\Aa_{L^T_{\omega} }))$~.
\Ee
\vskip 11pt

\item The real bisemisheaf of rings $ \phi _R(M_R(L_{\o v}))\otimes
\phi _L(M_L(L_{v})$ is a physical ``{\/\bf string field\/}'' having representations in the:
\Bena
\item bisemimodule $\phi _R(M^+_R(L_{\o v_+ }))\otimes \phi _L(M^+_L(L_{v_+ }))$ 
isomorphic to its toroidal equivalent\linebreak
$\phi _R(M^+_R(L^T_{\o v_+ }))\otimes \phi _L(M^+_L(L^T_{v_+ }))$~;
\item bisemimodule $\phi _R(M_R(\Aa_{L_{\o v} }))\otimes \phi _L(M_L(\Aa_{L_{v} }))$ 
isomorphic to its toroidal equivalent\linebreak
$\phi _R(M_R(\Aa_{L^T_{\o v} }))\otimes \phi _L(M_L(\Aa_{L^T_{v} }))$~.
\Ee
\vskip 11pt

\item ``String field'' representations are included into the corresponding brane field representations:
\Bi
\item $\phi _R(M^+_R(L_{\o v_+ }))\otimes \phi _L(M^+_L(L_{v_+ })) \subseteq
\phi _R(M^+_R(L_{\o\omega_+ }))\otimes \phi _L(M^+_L(L_{\omega_+ }))$~;

\item $\phi _R(M^+_R(L^T_{\o v_+ }))\otimes \phi _L(M^+_L(L^T_{v_+ }))\subseteq
\phi _R(M^+_R(L^T_{\o\omega_+ }))\otimes \phi _L(M^+_L(L^T_{\omega_+ }))$~;

\item $\phi _R(M_R(\Aa_{L_{\o v} }))\otimes \phi _L(M_L(\Aa_{L_{v} }))\subseteq
\phi _R(M_R(\Aa_{L_{\o\omega} }))\otimes \phi _L(M_L(\Aa_{L_{\omega} }))$~;

\item $\phi _R(M_R(\Aa_{L^T_{\o v} }))\otimes \phi _L(M_L(\Aa_{L^T_{v} }))\subseteq
\phi _R(M_R(\Aa_{L^T_{\o\omega} }))\otimes \phi _L(M_L(\Aa_{L^T_{\omega} }))$~.
\Ei
\Ee}
\vskip 11pt

\bpr
\Bean
\item The bisections $ \phi _R(M_{\o v_{\mu ,m_\mu }}) \otimes
\phi _L(M_{v_{\mu ,m_\mu }}) $ of the real bisemisheaf 
$\phi _R(M_R(L_{\o v }))\otimes \phi _L(M_L(L_{v })) $ are $\CC$-valued differentiable bifunctions on the conjugacy class representatives 
$ M_{\o v_{\mu ,m_\mu }} \otimes M_{v_{\mu ,m_\mu }} $ of the bilinear algebraic semigroup $\GL_2(L_{\o v} \times L_v)$~.  Now,  
$ M_{\o v_{\mu ,m_\mu }} \otimes M_{v_{\mu ,m_\mu }} $ is the (tensor) product of two symmetric (closed) strings at $\mu $ quanta in such a way that 
$ M_{\o v_{\mu ,m_\mu }} \otimes M_{v_{\mu ,m_\mu }} $ (and also 
$ \phi _R(M_{\o v_{\mu ,m_\mu }}) \otimes \phi _L(M_{v_{\mu ,m_\mu }}) $~) behaves like a harmonic oscillator \cite{Pie4}. So, the real bisemisheaf 
$ \phi _R(M_R(L_{\o v})) \otimes \phi _L(M_L(L_{v})) $~, constituted of a set
$ \Gamma (\phi _R(M_R(L_{\o v})) \otimes \phi _L(M_L(L_{v}))) $ of bisections, is a physical ``string field'' according to chapter 2, and, especially proposition 2.8.

This string field has two spin internal degrees of freedom, corresponding to the two possible directions of rotation of the strings, left and right symmetric strings rotating in opposite directions.
\vskip 11pt

\item As the real bisemisheaf $ \phi _R(M_R(L_{\o v})) \otimes \phi _L(M_L(L_{v})) $ is included into (and covers) the complex bisemisheaf
$ \phi _R(M_R(L_{\o \omega })) \otimes \phi _L(M_L(L_{\omega })) $ according to proposition 3.7 and as the complex conjugacy class representatives $M_{\omega _\mu }$ and $M_{\o\omega _\mu }$ over which are defined respectively the sections of the complex semisheaves $ \phi _R(M_R(L_{\o\omega })) $ and $\phi _L(M_L(L_{\omega })) $ are one-dimensional complex Lie semisubgroups, the complex bisemisheaf is a physical brane field.
\vskip 11pt

\item The brane field $ \phi _R(M_R(L_{\o \omega })) \otimes \phi _L(M_L(L_{\omega })) $~, and its toroidal equivalent 
$ \phi _R(M_R(L^T_{\o \omega })) \otimes \phi _L(M_L(L^T_{\omega })) $~, as well as the string field $ \phi _R(M_R(L_{\o v})) \otimes \phi _L(M_L(L_{v})) $~, and its toroidal equivalent $ \phi _R(M_R(L^T_{\o v})) \otimes \phi _L(M_L(L^T_{v})) $~, have the following representations given by the homomorphisms (described by arrows) in the commutative diagrams:
\Bi
\item $\begin{array}[t]{ccc}
 \phi _R(M_R(L^{(T)}_{\o \omega })) \otimes \phi _L(M_L(L^{(T)}_{\omega })) 
& \longrightarrow & 
 \phi _R(M^+_R(L^{(T)}_{\o \omega_+ })) \otimes \phi _L(M^+_L(L^{(T)}_{\omega_+ }))\\
\rotatebox{90}{\scalebox{1.5}{$\hookrightarrow $}} &&\rotatebox{90}{\scalebox{1.5}{$\hookrightarrow $}} \\
 \phi _R(M_R(L^{(T)}_{\o  v})) \otimes \phi _L(M_L(L^{(T)}_{v})) 
& \longrightarrow & 
 \phi _R(M^+_R(L^{(T)}_{\o v_+ })) \otimes \phi _L(M^+_L(L^{(T)}_{v_+ }))\end{array}$

\item $\begin{array}[t]{ccc}
 \phi _R(M_R(L^{(T)}_{\o \omega })) \otimes \phi _L(M_L(L^{(T)}_{\omega })) 
& \longrightarrow & 
 \phi _R(M_R(\Aa_{L^{(T)}_{\o \omega_+ }})) \otimes \phi _L(M_L(\Aa_{L^{(T)}_{\omega_+ }}))\\
\rotatebox{90}{\scalebox{1.5}{$\hookrightarrow $}} &&
\rotatebox{90}{\scalebox{1.5}{$\hookrightarrow $}} \\
 \phi _R(M_R(L^{(T)}_{\o  v})) \otimes \phi _L(M_L(L^{(T)}_{v})) 
& \longrightarrow & 
 \phi _R(M_R(\Aa_{L^{(T)}_{\o v_+ }})) 
\otimes \phi _L(M_L(\Aa_{L^{(T)}_{v_+ }}))\end{array}
\hfill \begin{array}[t]{c}\;\\[6pt]\;\\ \eop\end{array}$\Ei
\Ee
\vskip 11pt

\section[Real and complex smooth semivarieties]{Real and complex smooth semivarieties \cite{Mum}}

\Bi
\item A smooth linear general semivariety $\tau (M_L(L_{v}))$ (resp. $\tau (M_R(L_{\o v}))$~) is a modular representation semispace $M_L(L_{v})$ (resp. $M_R(L_{\o v})$~) composed of the family $\{M_{v_{\mu ,m_\mu }}\}$ (resp. $\{M_{\o v_{\mu ,m_\mu }}\}$~)
of disjoint real conjugacy class representatives together with a collection of charts from these real conjugacy class representatives to their complex equivalents:
\[ c_{\mu ,m_\mu }\ z^{\mu } : \quad M_{v_{\mu ,m_\mu }}
\longrightarrow M_{\omega _\mu } \qquad \text{(resp.} \quad 
c^*_{\mu ,m_\mu }\ z^{*\mu } : \quad M_{\o v_{\mu ,m_\mu }}
\longrightarrow M_{\o\omega _\mu } \ )\]
where:
\Bi
\item $z^\mu $ (resp. $z^{*\mu }$~) are coordinate functions on the corresponding conjugacy class representatives;
\item $c_{\mu ,m_\mu }$ (resp. $c^*_{\mu ,m_\mu }$~) are square roots of the eigenvalues of the $(\mu ,m_\mu )$-th coset representatives of the products, right by left, of Hecke operators \cite{Pie5}.
\Ei
\vskip 11pt

\item A smooth linear general semivariety $\tau (M_L(L_\omega ))$ (resp.
$\tau (M_R(L_{\o\omega }))$~) is  a modular representation semispace $M_L(L_\omega )$ (resp. $M_R(L_{\o\omega} )$~) composed of the family $\{M_{\omega _\mu }\}$ (resp.
$\{M_{\o\omega _\mu }\}$~) of disjoint complex conjugacy class representatives together with a collection of charts from these complex conjugacy class representatives to open sets in $\CC$~.
\Ei
\vskip 11pt

\section{Proposition}

{\em \Be
\item Let $\tau_h (M_L(L_v))$ (resp.
$\tau_h (M_R(L_{\o v}))$~) be a real compactified smooth general semivariety of which conjugacy class representatives $M_{v_{\mu ,m_\mu }}$ (resp. $M_{\o v_{\mu ,m_\mu }}$~) are glued together on a surface and on which regular functions:
\[ f_{v_{\mu ,m_\mu }}( z^{\mu }) : \quad M_{v_{\mu ,m_\mu }}
\longrightarrow F_{\omega _\mu } \qquad \qquad \text{(resp.} \quad 
f_{\o v_{\mu ,m_\mu }}( z^{*\mu }) : \quad M_{\o v_{\mu ,m_\mu }}
\longrightarrow F_{\o \omega  _\mu }  \ )\]
are considered.

Then, on this compactified semivariety $\tau_h (M_L(L_v))$ (resp.
$\tau_h (M_R(L_{\o v}))$~), the function $f_v(z)$ (resp. $f_{\o v}(z^*)$~), defined in a neighborhood of a point $z	_0$ (resp. $z^*_0$~) of $\CC$~, is holomorphic at $z_0$ (resp. $z^*_0$~) if we have the power series development:
\begin{alignat*}{3}
f^{(h)}_{v}(z) &= \sum_{\mu ,m_\mu } f_{v_{\mu ,m_\mu }} &=& \sum_{\mu ,m_\mu } c_{\mu ,m_\mu }(z-z_0)^\mu \\[11pt]
\text{(resp.} \quad 
f^{(h)}_{\o v}(z^*) &= \sum_{\mu ,m_\mu } f_{\o v_{\mu ,m_\mu }} 
&=& \sum_{\mu ,m_\mu } c^*_{\mu ,m_\mu }(z^*-z^*_0)^\mu \ ).
\end{alignat*}
\vskip 11pt

\item Let $\tau_h (M_L(L_\omega ))$ (resp.
$\tau_h (M_R(L_{\o \omega }))$~) denote the associated complex compactified smooth general semivariety of which conjugacy class representatives $M_{\omega _\mu }$ (resp. $M_{\o\omega _\mu }$~) are glued together on a surface and on which the regular functions:
\[ f_{\omega _{\mu}}( y^{\mu }) : \quad M_{\omega _{\mu}}
\longrightarrow F_{\omega _\mu } \qquad\qquad \text{(resp.} \quad 
f_{\o \omega _{\mu }}( y^{*\mu }) : \quad M_{\o\omega _{\mu}}
\longrightarrow F_{\o \omega  _\mu }  \ )\]
are defined.

Then, on this compactified semivariety $\tau_h (M_L(L_\omega ))$ (resp.
$\tau_h (M_R(L_{\o \omega }))$~), the function $f_\omega (y)$ (resp. $f_{\o\omega}(y^*)$~), defined in a neighborhood of a point $y_0$ (resp. $y^*_0$~) of $\CC$~, is holomorphic at $y_0$ (resp. $y^*_0$~) if we have the following power series development:
\begin{alignat*}{3}
f^{(h)}_{\omega }(y) &= \sum_{\mu } f_{\omega _{\mu}} &=& \sum_{\mu  } d_{\mu }(y-y_0)^\mu \\[11pt]
\text{(resp.} \quad 
f^{(h)}_{\o\omega }(y^*) &= \sum_{\mu } f_{\o\omega _{\mu}} &=& \sum_{\mu  } d^*_{\mu }(y^*-y^*_0)^\mu\ )
\end{alignat*}
where $d_\mu $ (resp. $d^*_\mu $~) are square roots of eigenvalues of coset representatives of products, right by left, of Hecke operators \cite{God}.
\Ee}
\vskip 11pt

\bpr
\Bi
\item This proposition presents a way of constructing a holomorpic function from functions on compactified conjugacy class representatives in such a way that each term of the power series development of the holomorphic function corresponds to a conjugacy class representative.

\item If the number of considered conjugacy class representatives tends to $\infty $ in the power series development, then it is hoped that this one is converging to $z$ (or to $y$~) in some neighborhood of $z_0$ (resp. $y_0$~) and is equal there to $f^{(h)}_{v}(z)$ (or to $f^{(h)}_{\omega }(y)$~).\epr
\Ei
\vskip 11pt

\section{Corollary}

{\em \Be
\item On the real smooth bisemivariety
$\tau_h (M_R(L_{\o v}) \otimes M_L(L_v))$ 
of which conjugacy class representatives  
$M_{\o v_{\mu ,m_\mu }} \otimes M_{v_{\mu ,m_\mu }}$ 
have been glued together, a bifunction 
$f^{(h)}_{\o v}(z^*)\otimes f^{(h)}_{v}(z)$~, 
defined in the neighborhood of a bipoint 
$(z^*_0\times z_0)$ of $\CC\times \CC$~, is holomorphic at 
$(z^*_0\times z_0)$ if there is the power series development:
\[f^{(h)}_{\o v}(z^*)\otimes f_v^{(h)}(z)
= \sum_{\mu ,m_\mu } c^*_{\mu ,m_\mu }\ c_{\mu ,m_\mu }
\ (z^*\ z-z^*_0\ z_0)^\mu  \;.\]

\item Similarly, on the complex smooth bisemivariety 
$\tau_h ( M_R(L_{\o \omega }) \otimes M_L(L_\omega ) )$ of which conjugacy class 
representatives  $M_{\o \omega _\mu } \otimes M_{\omega _\mu } $ have been 
glued together,  bifunction 
$f^{(h)}_{\o \omega }(y^*)\otimes f^{(h)}_{\omega }(y)$~, defined in the neighborhood of a bipoint $(y^*_0\times y_0)$ of $\CC\times \CC$~, is holomorphic at 
this bipoint if the have the following power series development:
\[f^{(h)}_{\o \omega }(y^*)\otimes f_\omega ^{(h)}(y)
= \sum_{\mu  } d^*_{\mu }\ d_{\mu }
\ (y^*\ y-y^*_0\ y_0)^\mu  \;.\]
\Ee}
\vskip 11pt

\bpr This is an adaptation of proposition 3.11 to the bilinear case.\epr
\vskip 11pt

\section{Polynomial functions on $2D$-semivarieties}
\Bi
\item If, instead of gluing the complex conjugacy class representatives
$M_{\omega _\mu}$ (resp. $M_{\o\omega _\mu}$~) on a surface, we stack them up in order to get a volume foliated by the two-dimensional conjugacy class representatives
$M_{\omega _\mu}$ (resp. $M_{\o\omega _\mu}$~), we shall obtain a three-dimensional compactified smooth semivariety $\tau _c(M_L(L_{\omega }))$ (resp.
$\tau _c(M_R(L_{\o\omega }))$~).

\item Similarly, as the set $\{M_{v_{\mu ,m_\mu }}\}_{m_\mu }$ (resp.
$\{M_{\o v_{\mu ,m_\mu }}\}_{\mu ,m_\mu }$~) of real conjugacy class representatives covers the surface $M_{\omega _\mu }$ (resp. $M_{\o\omega _\mu }$~) if they are glued together as it was done in proposition 3.11, the family  
$\{M_{v_{\mu ,m_\mu }}\}_{m_\mu }$ (resp.
$\{M_{\o v_{\mu ,m_\mu }}\}_{m_\mu }$~) of real conjugacy class representatives can be stacked up into a three-dimensional compactified smooth semivariety 
$\tau _c(M_L(L_{v}))$ (resp.
$\tau _c(M_R(L_{\o v}))$~) foliated by the set of two-dimensional compactified conjugacy class representatives 
$\{M_{v_{\mu ,m_\mu }}\}_{m_\mu }$ (resp.
$\{M_{\o v_{\mu ,m_\mu }}\}_{m_\mu }$~).
\vskip 11pt

\item So, a polynomial function on the smooth semivariety
 $\tau _c(M_L(L_{\omega }))$ (resp.
$\tau _c(M_R(L_{\o \omega }))$~) will be given by:
\[ f_{\omega }( y) =\sum_{\mu} d_\mu \ y^\mu 
 \qquad \text{(resp.} \quad 
f_{\o\omega }( y^*) =\sum_{\mu} d^*_\mu \ y^{*\mu} \ )\]
where $y^\mu $ (resp. $y^{*\mu }$~) are coordinate functions on the corresponding conjugacy class representatives.
\vskip 11pt

\item And, a polynomial function on the smooth semivariety 
$\tau _c(M_L(L_{v}))$ (resp.
$\tau _c(M_R(L_{\o v}))$~) will be given similarly by:
\[ f_{v}( z) =\sum_{\mu,m_\mu } c_{\mu ,m_\mu} \ z^\mu 
 \qquad \text{(resp.} \quad 
f_{\o v}( z^*) =\sum_{\mu,m_\mu } c^*_\mu \ z^{*\mu} \ ).\]
\Ei
\vskip 11pt

\section{Holomorphic and automorphic representations of brane and string fields}

\Bi
\item Sections 3.10 to 3.13 have introduced analytic (essentially holomorphic) representations of the brane and string fields by means of analytic representations respectively of the bisemimodules 
$\phi _R(M_R^+(L_{\o\omega _+})) \otimes \phi _L(M_L^+(L_{\omega _+}))$ and
$\phi _R(M_R^+(L_{\o v_+})) \otimes \phi _L(M_L^+(L_{v_+}))$ (see proposition 3.9).
\vskip 11pt

\item The two following next sections will deal with the corresponding toroidal analytic representations of these brane and string fields by means of the automorphic representations respectively of the bisemimodules
$\phi _R(M_R^+(L^T_{\o\omega _+})) \otimes \phi _L(M_L^+(L^T_{\omega _+}))$ and
$\phi _R(M_R^+(L^T_{\o v_+})) \otimes \phi _L(M_L^+(L^T_{v_+}))$~.
\Ei\vskip 11pt

\section{Proposition}

{\em
\Be 
\item An automorphic representation of the brane field bisemimodule 
$\phi _R(M_R^+(L^T_{\o\omega _+})) \otimes \phi _L(M_L^+(L^T_{\omega _+}))$ is given by the product, right by left, $ \Eis_R(2,\mu ) \otimes \Eis_L(2,\mu ) $ of the Fourier developments  of the normalized cusp forms of weight $k=2$~:
\begin{align*}
\Eis_L(2,\mu ) &\simeq \sum_\mu d'_\mu \ e^{2\pi i\mu z}\\
\Eis_R(2,\mu ) &\simeq \sum_\mu d^{*'}_\mu \ e^{-2\pi i\mu z}\;, \qquad z\in L_\omega \subset \CC\;.\end{align*}
\vskip 11pt

\item An automorphic representation of the string field bisemimodule 
$\phi _R(M_R^+(L^T_{\o v_+})) \otimes \phi _L(M_L^+(L^T_{v_+}))$ is given by the product, right by left, $ \Ellip_R(1,\mu,m_\mu  ) \otimes \Ellip_L(1,\mu,m_\mu  ) $ of global elliptic semimodules \cite{Pie4}
\begin{align*}
\Ellip_L(1,\mu,m_\mu  ) &=\sum_{\mu ,m_\mu } c'_{\mu,m_\mu } \ e^{2\pi i\mu x}\\
\Ellip_R(1,\mu,m_\mu  ) & = \sum_{\mu,m_\mu } c^{*'}_{\mu,m_\mu } \ e^{-2\pi i\mu z}\;, \qquad x\in L_v\subset \rit\;,\end{align*}
in such a way that
\[
\Ellip_L(1,\mu,m_\mu  ) \subseteq \Eis_L(2,\mu )\;, \qquad \qquad
\Ellip_R(1,\mu,m_\mu  ) \subseteq \Eis_R(2,\mu )\;.\]
\Ee}
\vskip 11pt

\bpr
\Be
\item According to section 3.4 and 3.8, the terms of the brane field bisemimodule
$\phi _R(M_R^+(L^T_{\o\omega _+})) \otimes \phi _L(M_L^+(L^T_{\omega _+}))$ are the bisections of the bisemisheaf $\phi _R(M_R(L^T_{\o\omega })) \otimes \phi _L(M_L(L^T_{\omega }))$~.  Now, these bisections are $\CC$-valued differentiable bifunctions on the conjugacy class representatives $M^T_{\o \omega _\mu }  \otimes M^T_{\omega _\mu }$ which are (tensor) products of right $2D$-(semi)tori $T^2_R[\mu ]$~, localized in the lower half space, by left $2D$-(semi)tori $T^2_L(\mu ]$~, localized in the upper half space \cite{Pie5}.

So, the analytic representation  of $M^T_{\o \omega _\mu }  \otimes M^T_{\omega _\mu }
= T^2_R[\mu ] \otimes T^2_L[\mu ]$ is given by the $\mu $-th term $d^{*'}_\mu \ e^{-2\pi i\mu z} \otimes d'_\mu \ e^{2\pi i\mu z}$ of $\Eis_R(2,\mu )\otimes \Eis_L(2,\mu )$~, leading to an automorphic representation of:
\Bi
\item the brane field bisemimodule $\phi _R(M^+_R(L^T_{\o\omega _+})) \otimes \phi _L(M^+_L(L^T_{\omega_+ }))$~,
\Ei
but also of
\Bi
\item the brane field $\phi _R(M_R(L_{\o\omega })) \otimes \phi _L(M_L(L_{\omega }))$~.  
\Ei
\vskip 11pt

\item Similarly, the terms of the string field bisemimodule 
$\phi _R(M_R^+(L^T_{\o v_+})) \otimes \phi _L(M_L^+(L^T_{v_+}))$ 
are the bisections of the real bisemisheaf 
$\phi _R(M_R(L^T_{\o v}) \otimes \phi _L(M_L(L^T_{v}))$~.  Now, these bisections are $\CC$-valued differentiable bifunctions on the conjugacy class representatives $\{M^T_{\o v_{\mu,m_\mu } }  \otimes M^T_{v_{\mu,m_\mu } }\}_{m_\mu }$ which are (tensor) products, right by left, $\{T^1_R[\mu,m_\mu  ] \otimes T^1_L(\mu ,m_\mu ]\}_{m_\mu }$  of (semi)circles and which cover their complex equivalents
$M^T_{\o\omega _\mu }\otimes M^T_{\omega _\mu }$~.

Thus, the analytic representation  of $M^T_{\o v_{\mu ,m_\mu }}  \otimes M^T_{v_{\mu ,m_\mu }}
= T^1_R[\mu,m_\mu  ] \otimes T^1_L[\mu,m_\mu  ]$ is given by the $(\mu,m_\mu ) $-th term $c^{*'}_{\mu,m_\mu } \ e^{-2\pi i\mu x} \otimes c'_{\mu,m_\mu } \ e^{2\pi i\mu x}$ of $\Ellip_R(1,\mu,m_\mu  )\otimes \Ellip_L(1,\mu,m_\mu  )$~, leading to an automorphic representation of:
\Bi
\item the string field bisemimodule $\phi _R(M^+_R(L^T_{\o v_+})) \otimes \phi _L(M^+_L(L^T_{v_+ }))$~,
\Ei
but also of
\Bi
\item the string field $\phi _R(M_R(L_{\o v})) \otimes \phi _L(M_L(L_{v}))$~. \epr 
\Ei
\Ee
\vskip 11pt

\section{Proposition}

{\em
\Be
\item An automorphic representation of the $\GL_2(\Aa_{L_{\o\omega }}\times \Aa_{L_\omega })$-bisemimodule $\phi _R(M_R(\Aa_{L_{\o\omega }})) \otimes \phi _L(M_L(\Aa_{L_\omega }))$ (and also of the bilinear algebraic semigroup
$\GL_2(\Aa_{L_{\o\omega }}\times \Aa_{L_\omega })$~) is given by
\[ \Eis_R(2,\mu )\otimes \Eis_L(2,\mu )
\simeq \sum_\mu (d^{*'}_\mu \ e^{-2\pi i\mu z}\otimes d_\mu \ e^{2\pi i\mu z})\;.\]
\vskip 11pt

\item An automorphic representation of the $\GL_2(\Aa_{L_{\o v}}\times \Aa_{L_v})$-bisemimodule $\phi _R(M_R(\Aa_{L_{\o v}})) \otimes \phi _L(M_L(\Aa_{L_v}))$ (and also of the bilinear algebraic semigroup
$\GL_2(\Aa_{L_{\o  v}}\times \Aa_{L_v})$~) is given by
\[ \Ellip_R(1,\mu,m_\mu  )\otimes \Ellip_L(1,\mu,m_\mu  )
\simeq \sum_{\mu,m_\mu } (c^{*'}_{\mu,m_\mu } \ e^{-2\pi i\mu x}\otimes c_{\mu,m_\mu } \ e^{2\pi i\mu x})\;.\]
\Ee}
\vskip 11pt

\bpr
\Be
\item Taking into account that:
\Bi
\item $\phi _R(M_R(\Aa_{L_{\o\omega }})) \otimes \phi _L(M_L(\Aa_{L_\omega }))=\prod_{\mu _p}( \phi _R(M_{\o\omega _{\mu _p}} )\otimes \phi _L(M_{\omega _{\mu _p}})) $~, where the direct product is taken over all the primary bisections of the complex bisemisheaf $ \phi _R(M_R(L_{\o\omega }))\otimes  \phi _L(M_R(L_{\omega }))$ according to section 3.8;

\item there is a homomorphism:
\[  \phi _R(M_R(L^{(T)}_{\o\omega }))\otimes  \phi _L(M _L(L^{(T)}_{\omega }))
\longrightarrow 
\phi _R(M_R(\Aa^{(T)}_{L_{\o\omega }})) \otimes \phi _L(M_L(\Aa^{(T)}_{L_\omega }))
\]
between the brane field 
$\phi _R(M_R(L_{\o\omega }))\otimes  \phi _L(M_L(L_{\omega }))$ and the bisemimodule
$\phi _R(M_R(\Aa_{L_{\o\omega }})) \otimes \phi _L(M_L(\Aa_{L_\omega }))$~, it becomes clear that $\Eis_R(2,\mu )\otimes \Eis_L(2,\mu )$ constitutes an automorphic representation of the bisemimodule 
$\phi _R(M_R(\Aa_{L_{\o\omega }})) \otimes \phi _L(M_L(\Aa_{L_\omega }))$~.
\Ei
\vskip 11pt

\item If we take into account:
\Bi
\item the development of the bisemimodule
$\phi _R(M_R(\Aa_{L_{\o v}})) \otimes \phi _L(M_L(\Aa_{L_v}))$ into
$\phi _R(M_R(\Aa_{L_{\o v}})) \otimes \phi _L(M_L(\Aa_{L_v}))=\prod_{\mu _p,m_{\mu _p}}( \phi _R(M_{\o v_{\mu _p,m_{\mu _p}}} )\otimes \phi _L(M_{v_{\mu _p,m_{\mu _p}}})) $~, according to section 3.8,

\item the commutative diagram:
\[ \begin{array}{cl}
\phi _R(M_R(L^{(T)}_{\o v})) \otimes \phi _L(M_L(L^{(T)}_v))
& \longrightarrow  \;\phi _R(M_R(\Aa_{L^{(T)}_{\o v}})) \otimes \phi _L(M_L(\Aa_{L^{(T)}_v})) \\
\scalebox{1.5}{$\downarrow $} & \; \scalebox{1.5}{$\nearrow $}\\
\multicolumn{2}{l}{
\quad\phi _R(M^+_R(L^{(T)}_{\o v})) \otimes \phi _L(M^+_L(L^{(T)}_v))}
\end{array}\]
with respect to proposition 3.9 (proof c)), it becomes clear that
$\Ellip_R(1,\mu ,m_\mu )\otimes\linebreak \Ellip_L(1,\mu ,m_\mu )$ 
constitutes an automorphic representation of the bisemimodule\linebreak
$ \phi _R(M_R(\Aa_{L_{\o v}})) \otimes \phi _L(M_L(\Aa_{L_v})) $~.\epr
\Ei
\Ee\vskip 11pt

\section{Proposition}

{\em
The brane field $\phi _R(M_R(L_{\o \omega })) \otimes \phi _L(M_L(L_\omega ))$ (as well as the string field $\phi _R(M_R(L_{\o v})) \otimes \phi _L(M_L(L_v))$~) is of solvable nature in the sense that:
\Bean
\item their bisections are embedded in the following sequence:
\[
\phi _R(M_{\o \omega_1 }) \otimes\phi _L(M_{\omega_1 }) 
\subseteq \cdots \subseteq
\phi _R(M_{\o \omega_\mu  }) \otimes\phi _L(M_{\omega_\mu  }) 
\subseteq \cdots \subseteq
\phi _R(M_{\o \omega_q}) \otimes\phi _L(M_{\omega_q}) \;, \quad \mu \le q\le \infty \;.\]

\item its representation given by the bisemimodule 
$\phi _R(M^+_R(L_{\o \omega_+ })) \otimes \phi _L(M^+_L(L_{\omega_+} ))$ (and its toroidal equivalent, see proposition 3.9) is such that it is generated in a solvable way by a tower of embedded subbisemimodules:
\begin{multline*}
\phi^{(1)} _R(M^+_R(L_{\o \omega_+ })) \otimes\phi^{(1)} _L(M^+_L(L_{\omega_+ })) 
\subseteq \cdots \subseteq
\phi^{(\mu )} _R(M^+_R(L_{\o \omega_+ })) \otimes\phi^{(\mu )} _L(M^+_L(L_{\omega_+ })) \\
\subseteq \cdots \subseteq
\phi^{(q)} _R(M^+_R(L_{\o \omega_+ })) \otimes\phi^{(q)} _L(M^+_L(L_{\omega_+ })) 
 \end{multline*}
where:
\Bi
\item $\phi^{(\mu )} _R(M^+_R(L_{\o \omega_+ })) 
\otimes\phi^{(\mu )} _L(M^+_L(L_{\omega_+ }) =\bigoplus_{\nu =1}^\mu (
\phi _R(M_{\o \omega_\nu }) \otimes\phi _L(M_{\omega_\nu }) )$~;

\item $\phi^{(q)} _R(M^+_R(L_{\o \omega_+ })) 
\otimes\phi^{(q)} _L(M^+_L(L_{\omega_+ }) \equiv
\phi _R(M^+_R(L_{\o \omega_+})) \otimes\phi _L(M^+_L(L_{\omega_+}) )$~.
\Ei

\item its holomorphic and automorphic representations
$ f^{(h)}_{\o\omega }(y^*) \otimes f^{(h)}_{\omega}(y) $ and
$\Eis_R(2,\mu )\otimes\Eis_L(2,\mu )$ are also generated in a solvable way.
\Ee}
\vskip 11pt

\bpr
\Bi
\item The brane field is of solvable nature because it is algebraic, that is to say, generated under the (bi)action of the product, right by left, of appropriate Galois or Weil groups \cite{Pie4}.

\item The holomorphic representation is said to be solvable if it is generated in a solvable way, i.e. that we have a tower of holomorphic subrepresentations given by:
\[
f^{(h)(1)}_{\o\omega }(y^*) \otimes f^{(h)(1)}_{\omega }(y) 
\subseteq \cdots \subseteq
f^{(h)(\mu )}_{\o\omega }(y^*) \otimes f^{(h)(\mu )}_{\omega }(y) 
\subseteq \cdots \subseteq
f^{(h)(q)}_{\o\omega }(y^*) \otimes f^{(h)(q)}_{\omega }(y) \]
where:
\Bi
\item 
$f^{(h)(\mu )}_{\o\omega }(y^*) \otimes f^{(h)(\mu )}_{\omega }(y) 
=\sum_{\nu =1}^\mu d^*_\nu \ d_\nu\ (y^*\ y-y^*_0\ y_0)^\nu $
\item $f^{(h)(q)}_{\o\omega }(y^*) \otimes f^{(h)(q)}_{\omega }(y) 
\equiv f^{(h)}_{\o\omega }(y^*) \otimes f^{(h)}_{\omega }(y)$~.
\Ei
\vskip 11pt

\item The automorphic representation $ \Eis_R(2,\mu ) \otimes \Eis_L(2,\mu )$ can be handled similarly.\epr
\Ei
\vskip 11pt

\section{Space-time fields of the vacua of (bisemi)fermions}

\Bean
\item Assume that the string field
$\phi _R(M_R(L_{\o v})) \otimes \phi _L(M_L(L_{v}))$~, included into the corresponding brane field $\phi _R(M_R(L_{\o \omega })) \otimes \phi _L(M_L(L_{\omega }))$~, has a representation as described in section 3.13, i.e. that the family
$\{ \phi _R(M_{\o v_{\mu ,m_\mu }}) \otimes \phi _L(M_{v_{\mu ,m_\mu }}) \}_{\mu ,m_\mu }$ of the sets
$\{ \phi _R(M_{\o v_{\mu ,m_\mu }}) \otimes \phi _L(M_{v_{\mu ,m_\mu }}) \}_{m_\mu }$ 
(~$m_\mu $ varying) of bisections, glued together into surfaces, is stacked up into a $3D$-volume.  

This string field then corresponds to a space field of the vacuum internal structure of an elementary fermion as described in section 2.7: it will be noted in condensed form $\widetilde M^S_{ST_R}\otimes \widetilde M^S_{ST_L}$~.
\vskip 11pt

\item Associated with this space field $\widetilde M^S_{ST_R}\otimes \widetilde M^S_{ST_L}$ of the vacuum, corresponds a time field
$\widetilde M^T_{ST_R}\otimes \widetilde M^T_{ST_L}$ of the vacuum which:
\Bi
\item is a string field 
$\phi _R(M_R(L_{\o v})) \otimes \phi _L(M_L(L_{v}))$ of which a family
$\{ \phi _R(M_{\o v_{\gamma ,m_\gamma }}) \otimes \phi _R(M_{v_{\gamma ,m_\gamma }}) \}_{\gamma ,m_\gamma }$ of bisections are not glued together and stacked up as for the corresponding space field.

So, this time field is one-dimensional.

\item is characterized by an internal algebraic dimension $\gamma $~, $1\le\gamma\le p\le\infty $ (see section 2.7 h)).

\item is related algebraically to the corresponding orthogonal space field by a $(\gamma _{r\to t}\circ \EE)$ morphism introduced in \cite{Pie6} and studied in \cite{Pie4}.
\Ei
\vskip 11pt

\item So, the space-time field of the internal structure of the vacuum of a bisemifermion is given by:
\[ \widetilde M^{TS}_{ST_R}\otimes \widetilde M^{TS}_{ST_L}
= (\widetilde M^{T}_{ST_R}\oplus \widetilde M^{S}_{ST_R})
\otimes (\widetilde M^{T}_{ST_L}\oplus \widetilde M^{S}_{ST_L})\;.\]
It can undergo a blowup isomorphism decomposing it into a diagonal structure field and into off-diagonal magnetic and electric interaction fields as developed in the next proposition.
\Ee
\vskip 11pt

\section{Proposition}

{\em
The 10-dimensional space time field $\widetilde M^{TS}_{ST_R}\otimes \widetilde M^{TS}_{ST_L}$ can be transformed under the blowup isomorphism
\[ S_L : \quad
\widetilde M^{TS}_{ST_R}\otimes \widetilde M^{TS}_{ST_L}
\longrightarrow 
(\widetilde M^{TS}_{ST_R}\otimes_D \widetilde M^{TS}_{ST_L})\oplus
( \widetilde M^{S}_{ST_R}\otimes_{\rm magn} \widetilde M^{S}_{ST_L})
\oplus
(\widetilde M^{S-(T)}_{ST_R}\otimes_{\rm elec} \widetilde M^{S-(T)}_{ST_L})\]
into the following disconnected fields:
\Bean
\item a diagonal field
$(\widetilde M^{TS}_{ST_R}\otimes_D \widetilde M^{TS}_{ST_L})$ of dimension 
4 characterized by a diagonal orthogonal $4D$-basis $\{e^\alpha \otimes f_\alpha \}^3_{\alpha =0}$~, $\forall\ e^\alpha \in 
\widetilde M^{TS}_{ST_R} $ and $f_\alpha \in \widetilde M^{TS}_{ST_L}$~.

\item a magnetic field
$(\widetilde M^{S}_{ST_R}\otimes_{\rm magn} \widetilde M^S_{ST_L})$ characterized by a $3D$-non  orthogonal basis $(e^\alpha \otimes f_\beta )^3_{\alpha \neq \beta =1}$~.

\item an electric field 
$(\widetilde M^{S}_{ST_R}\otimes_{\rm elec} \widetilde M^{T}_{ST_L})$ or
$(\widetilde M^{T}_{ST_R}\otimes_{\rm elec} \widetilde M^{S}_{ST_L})$ characterized by a $3D$-non orthogonal basis.
\Ee
}
\vskip 11pt

\bpr
\Be
\item First, let us remark that the time and space diagonal fields 
$(\widetilde M^{T}_{ST_R}\otimes_D \widetilde M^{T}_{ST_L})$ and
$(\widetilde M^{S}_{ST_R}\otimes_D \widetilde M^{S}_{ST_L})$ are the string fields
$\phi _R(M_R(L_{\o v}))\otimes \phi _L(M_L(L_v))$ studied until now in this chapter, the complete tensor product ``~$\otimes$~'' corresponding to the diagonal tensor product
``~$\otimes_D$~'' since the bisections were not necessarily considered as compactified.
\vskip 11pt

\item The blowup $S_L$ was introduced in chapter 1 of \cite{Pie4}.  It corresponds to the following decomposition starting from section 3.18 c):
\begin{align*}
\widetilde M^{TS}_{ST_R}\otimes \widetilde M^{TS}_{ST_L}
&= (\widetilde M^{T}_{ST_R}\oplus \widetilde M^{S}_{ST_R}) \otimes
(\widetilde M^{T}_{ST_L}\oplus \widetilde M^{S}_{ST_L})\\
&= (\widetilde M^{T}_{ST_R}\otimes \widetilde M^{T}_{ST_L})
\oplus (\widetilde M^{S}_{ST_R}\otimes \widetilde M^{S}_{ST_L})
\oplus (\widetilde M^{T}_{ST_R}\otimes_{\rm elec} \widetilde M^{S}_{ST_L})
\oplus (\widetilde M^{S}_{ST_R}\otimes_{\rm elec} \widetilde M^{T}_{ST_L})
\end{align*}
in such a way that:
\[(\widetilde M^{T}_{ST_R}\otimes \widetilde M^{T}_{ST_L})\oplus
(\widetilde M^{S}_{ST_R}\otimes \widetilde M^{S}_{ST_L})
= (\widetilde M^{TS}_{ST_R}\otimes_D \widetilde M^{TS}_{ST_L})
\oplus (\widetilde M^{S}_{ST_R}\otimes_{\rm magn} \widetilde M^{S}_{ST_L})\]
where:
\Bi
\item $(\widetilde M^{TS}_{ST_R}\otimes_D \widetilde M^{TS}_{ST_L})
= (\widetilde M^{T}_{ST_R}\otimes_D \widetilde M^{T}_{ST_L})
\oplus (\widetilde M^{S}_{ST_R}\otimes_D \widetilde M^{S}_{ST_L})$
denotes the space-time field of the internal vacuum structure of a bisemifermion; this space-time field is given by a diagonal tensor product between the right space-time semisheaf
$\widetilde M^{TS}_{ST_R}$ and is left equivalent $\widetilde M^{TS}_{ST_L}$~.

\item the magnetic bisemisheaf
$(\widetilde M^{S}_{ST_R}\otimes_{\rm magn} \widetilde M^{S}_{ST_L})$ results from the off-diagonal interactions between the space field
$(\widetilde M^{S}_{ST_R}\otimes_{(D)} \widetilde M^{S}_{ST_L})$ as described in chapter 1 of \cite{Pie4}.
\Ei
\vskip 11pt

Finally, the electric field
$(\widetilde M^{T}_{ST_R}\otimes_{\rm elec} \widetilde M^{S}_{ST_L})$ or
$(\widetilde M^{S}_{ST_R}\otimes_{\rm elec} \widetilde M^{T}_{ST_L})$ results from (off-diagonal) interactions between the right part of time (or space) semisheaf and left part of the space (or time) semisheaf.\epr
\Ee
\vskip 11pt

\section{Algebraic bilinear Hilbert spaces}

\Bean
\item An algebraic left or right {\bbf extended (internal) bilinear Hilbert space $H^+_a$ or $H^-_a$\/}~, introduced in \cite{Pie1}, can be obtained from the complete bisemisheaf
$( \widetilde M^{TS}_{ST_R}\otimes \widetilde M^{TS}_{ST_L} )$ by considering a map
\begin{align*}
B_L\circ p_L : \qquad &\widetilde M^{TS}_{ST_R}\otimes \widetilde M^{TS}_{ST_L}
\longrightarrow H^+_a
= \widetilde M^{TS}_{ST_{L_R}}\otimes \widetilde M^{TS}_{ST_L} 
\\
\text{or} \qquad
B_R\circ p_R : \qquad &\widetilde M^{TS}_{ST_R}\otimes \widetilde M^{TS}_{ST_L}
\longrightarrow H^-_a
= \widetilde M^{TS}_{ST_{R_L}}\otimes \widetilde M^{TS}_{ST_R} 
\;,\end{align*}
where, according to chapter 3 of \cite{Pie4},
\Bi
\item $p_L$ (resp. $p_R$~) is a projective linear map from
$\widetilde M^{TS}_{ST_R}$~, noted $\widetilde M^{TS}_{ST_{L_R}} $
(resp. $\widetilde M^{TS}_{ST_L}$~, noted $\widetilde M^{TS}_{ST_{R_L}} $~) into
$\widetilde M^{TS}_{ST_L}$ 
(resp. $\widetilde M^{TS}_{ST_R}$~);

\item $B_L$ (resp. $B_R$~) is a bijective linear isometric map;
\Ei
and a complete internal bilinear form on $H^+_a$ and $H^-_a$~.
\vskip 11pt

\item \Bi
\item Similarly, an algebraic left or right {\bbf internal bilinear (diagonal) Hilbert space $\Hs^+_a$ or $\Hs^-_a$\/} will be obtained as follows:
\[ 
\widetilde M^{TS}_{ST_R}\otimes_D \widetilde M^{TS}_{ST_L}\quad
{\renewcommand{\arraystretch}{.4}
\begin{array}{c}
\rotatebox{10}{$\xrightarrow{\;{\scriptstyle B_L\circ p_L}\;}$} \\
\rotatebox{-10}{$\xrightarrow[\;{\scriptstyle B_R\circ p_R}\;]{} $} \end{array}} \quad
\begin{array}{c}
\Hs^+_a
= \widetilde M^{TS}_{ST_{L_R}}\otimes_D \widetilde M^{TS}_{ST_L} 
 \\ \Hs^-_a
= \widetilde M^{TS}_{ST_{R_L}}\otimes_D \widetilde M^{TS}_{ST_R} 
\end{array}\]

\item An algebraic left or right {\bbf internal bilinear magnetic space 
${\rm v}^+_{m;a}$ or ${\rm v}^-_{m;a}$\/} will be obtained by taking into account:
\[ 
\widetilde M^{S}_{ST_R}\otimes_{\rm magn} \widetilde M^{S}_{ST_L}\quad
{\renewcommand{\arraystretch}{.4}
\begin{array}{c}
\rotatebox{10}{$\xrightarrow{\;{\scriptstyle B_L\circ p_L}\;}$} \\
\rotatebox{-10}{$\xrightarrow[\;{\scriptstyle B_R\circ p_R}\;]{} $} \end{array}} \quad
\begin{array}{c}
{\rm v}^+_{m;a}
= \widetilde M^{TS}_{ST_{L_R}}\otimes_{\rm magn} \widetilde M^{TS}_{ST_L} 
 \\ {\rm v}^-_{m;a}
= \widetilde M^{TS}_{ST_{R_L}}\otimes_{\rm magn} \widetilde M^{TS}_{ST_R} 
\end{array}\]

\item And an algebraic left or right {\bbf internal bilinear electric space 
${\rm v}^+_{e;a}$ or ${\rm v}^-_{e;a}$\/} will be obtained by considering:
\[ 
\widetilde M^{S}_{ST_R}\otimes_{\rm elec} \widetilde M^{T}_{ST_L}\quad
{\renewcommand{\arraystretch}{.4}
\begin{array}{c}
\rotatebox{10}{$\xrightarrow{\;{\scriptstyle B_L\circ p_L}\;}$} \\
\rotatebox{-10}{$\xrightarrow[\;{\scriptstyle B_R\circ p_R}\;]{} $} \end{array}} \quad
\begin{array}{c}
{\rm v}^+_{e;a} 
= \widetilde M^{TS}_{ST_{L_R}}\otimes_{\rm elec} \widetilde M^{TS}_{ST_L} 
\\ {\rm v}^-_{e;a}
= \widetilde M^{TS}_{ST_{R_L}}\otimes_{\rm elec} \widetilde M^{TS}_{ST_R} 
\end{array}\]
\Ei

Furthermore, it is assumed that these bilinear spaces are endowed with the corresponding internal bilinear forms.
\vskip 11pt

\item \Bi
\item The {\bf bielements\/} of the bilinear (diagonal) Hilbert spaces 
$\Hs^+_a$ and $\Hs^-_a$ are diagonal products of corresponding right and left sections as considered in section 3.18 with the suitable maps $B_L\circ p_L$ or $B_R\circ p_R$~.

\item The bielements of the bilinear magnetic spaces ${\rm v}^+_{m;a}$ or ${\rm v}^-_{m;a}$ are magnetic products (~$\times_{\rm magn}$~), in the sense of proposition 3.19, of right and left space sections ``pulled out'' from the {\bf extended\/} bilinear Hilbert spaces $H^+_a$ or $H^-_a$ by a magnetic biendomorphism $(E_R\otimes_{\rm magn}E_L)$ based on Galois antibiautomorphisms as developed in \cite{Pie4}.

\item Similarly, the bielements of the bilinear electric spaces ${\rm v}^+_{e;a}$ or ${\rm v}^-_{e;a}$ are electric products (~$\times_{\rm elec}$~) of right and left space (or vice-versa) sections ``pulled out'' from the {extended\/} bilinear Hilbert spaces $H^+_a$ or $H^-_a$~.
\Ei
\Ee
\vskip 11pt

\section{Introducing chapter 4}

The vacuum fields considered in this chapter are vacuum {\bf ``classical'' fields\/} \cite{Wig} with respect to the terminology of QFT.  The corresponding operator valued fields will be considered in the next chapter.
\vskip 11pt

\chapter{States of the vacuum and mass string fields of (bisemi)fermions}

\thispagestyle{empty}

\section{States of the space-time string field of the vacuum}

\subsection{Bialgebras of von Neumann}

\Bi
\item Let $H^{\pm}_a$ denote a \lr extended internal bilinear Hilbert space and let $\Hs^{\pm}_a$ be the corresponding \lr diagonal internal bilinear Hilbert space characterized by an orthonormal basis.
\vskip 11pt

\item 	A bialgebra of von Neumann $\MM^a\RxL (H^{\pm}_a)$ on the extended bilinear Hilbert space $H^{\pm}_a$ is an involutive subbialgebra of bounded operators on $H^{\pm}_a$ having a closed norm topology.

Similarly, a bialgebra of von Neumann $\MM^a\RxL (\Hs^{\pm}_a)$ on the diagonal bilinear Hilbert space $\Hs^{\pm}_a$ is an involutive subbialgebra of bounded operators on $\Hs^{\pm}_a$ having a closed norm topology.
\vskip 11pt

\item \begin{tabbing}
Let \quad \= $H^+_a \simeq \widetilde M^{TS}_{ST_{L_R}} \otimes \widetilde M^{TS}_{ST_{L}} $ \quad \= (resp. \quad 
$H^-_a \simeq \widetilde M^{TS}_{ST_{R_L}} \otimes \widetilde M^{TS}_{ST_{R}} $~)\\
and \quad \> $\Hs^+_a \simeq \widetilde M^{TS}_{ST_{L_R}} \otimes_D \widetilde M^{TS}_{ST_{L}} $ \quad \> (resp. \quad 
$\Hs^-_a \simeq \widetilde M^{TS}_{ST_{R_L}} \otimes_D \widetilde M^{TS}_{ST_{R}} $~)
\end{tabbing}

be the extended and diagonal bilinear Hilbert spaces as constructed on bisemisheaves according to section 3.20 and chapter 3 of \cite{Pie4}.

Let $(T_R\otimes T_L)$ be the tensor product of the right and left differential operators $T_R$ and $T_L$ acting respectively on the semisheaves $\widetilde M^{TS}_{ST_{L_R}}$ and $\widetilde M^{TS}_{ST_L}$ of $H^+_a$ in such a way that
$(T_R\otimes T_L)\in M^a\RxL(H^+_a)$~.
\Ei
\vskip 11pt

\subsection{Proposition}

{\em The action of the differential bioperator $T_R\otimes T_L$ on the extended bilinear Hilbert space $H^+_a$~:
\Be
\item consists in mapping the bisemisheaf 
$\widetilde M^{TS}_{ST_{L_R}} \otimes \widetilde M^{TS}_{ST_{L}} \subset H^+_a$ into the corresponding 
bisemisheaf
$\widetilde M^{TS_p}_{ST_{L_R}} \otimes \widetilde M^{TS_p}_{ST_{L}} $ which is shifted into  its geometrical dimensions onto its algebraic dimensions
\[ T_R\otimes T_L:\quad
\widetilde M^{TS}_{ST_{L_R}} \otimes \widetilde M^{TS}_{ST_{L}} 
\longrightarrow 
\widetilde M^{TS_p}_{ST_{L_R}} \otimes \widetilde M^{TS_p}_{ST_{L}} \;.\]

\item is associated with the generation of the tangent bibundle $\TAN
( \widetilde M^{TS}_{ST_{L_R}} \otimes \widetilde M^{TS}_{ST_{L}} )$ whose total space is the shifted  bisemisheaf $\widetilde M^{TS_p}_{ST_{L_R}} \otimes \widetilde M^{TS_p}_{ST_{L}} $ which is an {\bf operator valued string field\/} according to QFT.
\Ee}
\vskip 11pt

\bpr
\Bi
\item The bioperator $T_R\otimes T_L$ maps the bisemisheaf
$\widetilde M^{TS}_{ST_{L_R}} \otimes \widetilde M^{TS}_{ST_{L}} $ into its shifted  equivalent
$M^{TS_p}_{ST_{L_R}} \otimes \widetilde M^{TS_p}_{ST_{L}} $ since this latter belongs to the derived category of string fields $\phi _{L_R}(M_{L_R}(L_{\o v}))
\otimes \phi _L(M_L(L_v))$ shifted in the four geometrical space-time dimensions of
$M_{L_R}(L_{\o v})$ and of $M_L(L_v)$~.
\vskip 11pt

\item On the other hand, 
$ \widetilde M^{TS}_{ST_{L_R}} \otimes \widetilde M^{TS}_{ST_{L}} $ decomposes, according to section 3.18, into:
\[
\widetilde M^{TS}_{ST_{L_R}} \otimes \widetilde M^{TS}_{ST_{L}} 
= ( \widetilde M^{T}_{ST_{L_R}} \oplus \widetilde M^{S}_{ST_{L_R}} )\otimes
( \widetilde M^{T}_{ST_{L}} \oplus \widetilde M^{S}_{ST_{L}} )\;.\]

Now, the time semisheaf
$\widetilde M^{T}_{ST_{L}} $ (resp. $\widetilde M^{T}_{ST_{L_R}} $~) is characterized by the set of its $p$ sections\linebreak $\{\widetilde M^T_{v_{\gamma ,m_\gamma }}\}_{\gamma =1,m_\gamma }^p$ (resp. $\{\widetilde M^T_{\o v_{\gamma ,m_\gamma }}\}_{\gamma =1,m_\gamma }^p$~)
having multiplicities $m^{(\gamma )}=\sup (m_\gamma)+1 $~.

While the space semisheaf $\widetilde M^{S}_{ST_{L}} $ (resp. $\widetilde M^{S}_{ST_{L_R}} $~) is characterized by the set of its $q$ sections $\{\widetilde M^S_{v_{\mu ,m_\mu }}\}^q_{\mu =1,m_\mu }$ (resp.
$\{\widetilde M^S_{\o v_{\mu ,m_\mu }}\}^q_{\mu =1,m_\mu }$~) having multiplicities $m^{(\mu )}=\sup (m_\mu) +1$~.

So, the number of algebraic dimensions of time is $p$ and the number of algebraic dimensions of space is $q$~.
\vskip 11pt

\item Then, the action of the differentiable operator $T_L$ (resp. $T_R$~) on the semisheaves
$(\widetilde M^{T}_{ST_{L}}\oplus \widetilde M^{S}_{ST_{L}}) $
(resp. $(\widetilde M^{T}_{ST_{L_R}}\oplus \widetilde M^{S}_{ST_{L_R}}) $~) splits into
\[ T_L=T_L^T+T_L^S \qquad \text{(resp.} \quad T_R=T_R^T+T_R^S\ )\]
in such a way that $T_L^T$ (resp. $T_R^T$~) operates on
$( \widetilde M^{T}_{ST_{L}} $ (resp. $\widetilde M^{T}_{ST_{L_R}}) $~)
and $T_L^S$ (resp. $T_R^S$~) operates on
$\widetilde M^{S}_{ST_{L}} $ (resp. $\widetilde M^{S}_{ST_{L_R}} $~).
\vskip 11pt

\item Furthermore, if we take into account the structure of the semisheaves with respect to their sections, the operator
$T_L$ (resp. $T_R$~) decomposes, as a random operator, into:
\begin{align*}
T_L &= \{T^T_L(\gamma )+T^S_L(\mu )\}^p_{\gamma =1,}\;^q_{\mu =1}\\
\text{(resp.} \quad
T_R &= \{T^T_R(\gamma )+T^S_R(\mu )\}^p_{\gamma =1,}\;^q_{\mu =1}\ )
\end{align*}
following a set of operators corresponding to the algebraic dimensions.

Thus, the biaction of $T_R\otimes T_L$
\[T_R\otimes T_L : \quad 
\widetilde M^{TS}_{ST_{L_R}}\otimes\widetilde M^{TS}_{ST_{L}}
\longrightarrow 
\widetilde M^{TS_p}_{ST_{L_R}}\otimes\widetilde M^{TS_p}_{ST_{L}}\]
decomposes into the set of biactions
\begin{multline*}
\big\{ (T^T_R(\gamma )+T_R^S(\mu )) \otimes (T^T_L(\gamma )+T_L^S(\mu )) :\\
\L( 
\L\{ \widetilde M^T_{\o v_{\gamma ,m_\gamma }}
\R\}_{m_\gamma } +
\L\{
\widetilde M^S_{\o v_{\mu ,m_\mu }}
\R\}_{m_\mu } 
\R) \otimes
\L(
\L\{ \widetilde M^T_{v_{\gamma ,m_\gamma }}
\R\}_{m_\gamma } +
\L\{ \widetilde M^S_{v_{\mu ,m_\mu }}
\R\}_{m_\mu } 
\R)_{\gamma ,\mu }\\
\longrightarrow 
\L( \L\{ \widetilde M^{T_p}_{\o v_{\gamma ,m_\gamma }}
\R\}_{m_\gamma } +
\L\{ \widetilde M^{S_p}_{\o v_{\mu ,m_\mu }}
\R\}_{m_\mu } 
\R)\otimes
\L( \L\{ \widetilde M^{T_p}_{v_{\gamma ,m_\gamma }}
\R\}_{m_\gamma } +
\L\{ \widetilde M^{S_p}_{v_{\mu ,m_\mu }}
\R\}_{m_\mu } 
\R)_{\gamma ,\mu }
\end{multline*}
on the bisections 
$\L\{\widetilde M^T_{\o v_{\gamma ,m_\gamma }} \otimes
\widetilde M^T_{v_{\gamma ,m_\gamma }}\R\}_{m_\gamma }$~, \ldots, and so on, into their shifted  equivalents\linebreak
$\L\{\widetilde M^{T_p}_{\o v_{\gamma ,m_\gamma }} \otimes
\widetilde M^{T_p}_{v_{\gamma ,m_\gamma }}\R\}_{m_\gamma }$~.
\vskip 11pt

\item The tangent bibundle $\TAN ( \widetilde M^{TS}_{ST_{L_R}} \otimes
\widetilde M^{TS}_{ST_{L}} )$ is characterized by:
\Bi
\item its base given by the bisemisheaf 
$\widetilde M^{TS}_{ST_{L_R}} \otimes
\widetilde M^{TS}_{ST_{L}} $~;
\item its total space given by the corresponding shifted  bisemisheaf
$\widetilde M^{TS_p}_{ST_{L_R}} \otimes
\widetilde M^{TS_p}_{ST_{L}} $~.
\item its projective map given by $(T_R^{-1}\otimes T_L^{-1})$~.\epr
\Ei
\Ei
\vskip 11pt

\subsection{Proposition}

{\em Let $r=p+q$ be the number of algebraic dimensions of time and space.

Then, as a consequence of the solvability of the extended bilinear Hilbert space $H^+_a$ of space-time, a tower of modular subbialgebras of von Neumann can be defined.
}
\vskip 11pt

\bpr
\Bi
\item As the bisemisheaves 
$\widetilde M^{TS}_{ST_{L_R}} \otimes
\widetilde M^{TS}_{ST_{L}} $ of $H^+_a$ are defined over algebraic bilinear semigroups (see section 3.6), their bisections on the conjugacy classes of these algebraic bilinear semigroups correspond to extended bilinear Hilbert subspaces which form the following sequence of embedded subspaces:
\[H^+_a(1) \subset \cdots \subset
H^+_a(\gamma ) \subset \cdots \subset
H^+_a(\sigma ) \subset \cdots \subset
H^+_a(r)\;, \qquad 1\le \sigma \le r\;, \]
where $\sigma $ denotes a general algebraic dimension ``covering'' the running indices $\gamma$ and $\mu $ respectively of time and space.

So, $H^+_a$ will be said to be solvable by extending this concept from solvable groups.
\vskip 11pt

\item As a consequence, the bialgebra of von Neumann $ \MM^a\RxL(H^+_a)$ on $H^+_a$ also decomposes according to a sequence of corresponding embedded subbialgebras:
\be \MM^a\RxL(H^+_a(1)) \subset \cdots \subset
 \MM^a\RxL(H^+_a(\sigma )) \subset \cdots \subset
 \MM^a\RxL(H^+_a(r)) \;.\tag*{\eop}\ee
\Ei
\vskip 11pt

\subsection{Tower of sums of extended bilinear Hilbert subspaces}

Taking into account the representation of the string field into the sum of its bisections according to section 3.8 and proposition 3.9, we can introduce a tower of embedded extended bilinear Hilbert subspaces:
\[ H^+_a\{1\} \subset \cdots \subset
H^+_a\{\sigma \} \subset \cdots \subset
H^+_a\{r\} \]
in such a way that:
\Bi
\item $H^+_a\{\sigma \}=\bigoplus_{\tau =1}^\sigma H^+_a(\tau _+)$~, \quad 
where $\begin{array}[t]{ll}
H^+_a(\tau _+) &= \bigoplus_{m_\tau }H^+_a(\tau ,m_\tau )\\
&\simeq \bigoplus_{m_\tau } \widetilde M^{TS}_{\o v_{\tau ,m_\tau }} \otimes
\widetilde M^{TS}_{v_{\tau ,m_\tau }} \end{array}$

denotes an extended bilinear Hilbert subspace characterized by the sum over the multiples of $H^+_a(\tau )$~.

 \item $H^+_a\{r\}=\bigoplus_{\tau =1}^r H^+_a(\tau _+)$~.

\item $H^+_a\{1\} \equiv H^+_a(1_+)$~.
\Ei

So, every extended bilinear Hilbert subspace $H^+_a\{\sigma \}$~, $1\le \sigma \le r$~, is the sum of the extended bilinear Hilbert subspaces $H^+_a(\tau _+)$~, the index $\tau $ running over the algebraic dimensions inferior to it, in such a way that the Hilbert subspace $H^+_a(\tau _+)$ is summed over its multiples $H^+_a(\tau ,m_\tau )$~.
\vskip 11pt

\subsection{Shifted  solvable bilinear Hilbert spaces}

\Bi
\item To the shifted bisemisheaf
$ \widetilde M^{TS_p}_{ST_{L_R}}\otimes 
\widetilde M^{TS_p}_{ST_{L}}$ corresponds a shifted  extended bilinear Hilbert space $H^+_{ap}$ and to its diagonal equivalent 
$ \widetilde M^{TS_p}_{ST_{L_R}}\otimes_D 
\widetilde M^{TS_p}_{ST_{L}}$ corresponds a shifted  diagonal bilinear Hilbert space $\Hs^+_{ap}$ in such a way that 
\[ H^+_{ap}\simeq \widetilde M^{TS_p}_{ST_{L_R}} \otimes \widetilde M^{TS_p}_{ST_{L}} 
\;, \qquad 
\Hs^+_{ap}\simeq \widetilde M^{TS_p}_{ST_{L_R}} \otimes_D \widetilde M^{TS_p}_{ST_{L}} 
\;.\]
\vskip 11pt

\item As a consequence of propositions 4.1.2 and 4.1.3, $H^+_{ap}$ is solvable.  So, we have a sequence of embedded shifted  extended bilinear Hilbert subspaces:
\[ H^+_{ap}(1) \subset \cdots \subset
H^+_{ap}(\sigma ) \subset \cdots \subset
H^+_{ap}(r) \]
and a tower of sums of shifted  extended bilinear Hilbert subspaces:
\[ H^+_{ap}\{1\} \subset \cdots \subset
H^+_{ap}\{\sigma \} \subset \cdots \subset
H^+_{ap}\{r\} \]
which can be defined as in section 4.4 by:
\[H^+_{ap}\{\sigma \} = \bigoplus_{\tau =1}^\sigma  H^+_{ap}(\tau _+) \]
where $H^+_{ap}(\tau _+) = \bigoplus_{m_\tau }H^+_{ap}(\tau ,m_\tau )$~.
\Ei
\vskip 11pt

\subsection{Projectors and space-time states of fields}

\Bi
\item As a consequence of the solvability of the extended bilinear Hilbert space $H^+_a$ and of its shifted  equivalent $H^+_{ap}$~, we can introduce respectively on these the set of (bi)projectors $P^a\RxL\{\sigma \}$ and $P^{ap}\RxL\{\sigma \}$ by the mappings:
\begin{align*}
P^a\RxL \{\sigma \} &: \quad H^+_a \longrightarrow H^+_a\{\sigma \}\;, \quad \forall\ \sigma \ , \; 1\le \sigma \le r\;, \\
P^{ap}\RxL \{\sigma \} &: \quad H^+_{a,p} \longrightarrow H^+_{a,p}\{\sigma \} \;.\end{align*}
\vskip 11pt

\item Similarly, (bi)projectors $P^a\RxDL\{\sigma \}$ on the solvable diagonal bilinear Hilbert space $\Hs^+_a$ can be introduced 
by the mappings:
\[
P^a\RxDL \{\sigma \} : \quad \Hs^+_a \longrightarrow \Hs^+_a\{\sigma \}\;, \quad \forall\ \sigma \ , \; 1\le \sigma \le r\;, \]
in such a way that the $\GL_2(L_{\o v_+}\times L_{v_+})$-bisemimodule
$
\widetilde M^{TS_+}_{ST_{L_R}} \otimes_D \widetilde M^{TS_+}_{ST_{L}} 
= \bigoplus_{\tau =1}^r \bigoplus_{m_\tau }
(\widetilde M^{TS}_{ \o v_{\tau ,m_\tau }} \otimes_D \widetilde M^{TS}_{ v_{\tau ,m_\tau }} )$ of $\Hs^+_a$ be sent into the $\sigma $-th $\GL_2(L_{\o v}\times L_v)$-subbisemimodule
$
\widetilde M^{TS_+}_{ST_{L_R}}\{\sigma \} \otimes_D \widetilde M^{TS_+}_{ST_{L}} \{\sigma \}
= \bigoplus_{\tau =1}^\sigma  \bigoplus_{m_\tau }
( \widetilde M^{TS}_{ \o v_{\tau ,m_\tau }} \otimes_D \widetilde M^{TS}_{ v_{\tau ,m_\tau }} )$ of $\Hs^+_a\{\sigma \}$~.
\vskip 11pt

\item This subbisemimodule $\widetilde M^{TS}_{ ST_{L_R}} \{\sigma \} \otimes_D \widetilde M^{TS}_{ ST_L}\{\sigma \} $
is the {\bbf $\sigma $-th (bi)state of the space-time field\/} 
$\widetilde M^{TS}_{ ST_{L_R} } \otimes_D \widetilde M^{TS}_{ ST_L } $ if it is an eigen(bi)state of an eigen(bi)value as described in the following.
\Ei
\vskip 11pt

\subsection{Towers of sums of von Neumann subbialgebras}

\Bi
\item As the operator $T_L$ (resp. $T_R$~) was introduced in proposition 4.2 as decomposing into a set of random operators in accordance with the conjugacy classes of (the bisemisheaf on) the algebraic bilinear semigroup on which $T_L$ (resp. $T_R$~) operates, {\bf sums of products, right by left, of random operators\/} can be generated as follows:
\[ T\RxL\{\sigma \} = \bigoplus_{\tau =1}^\sigma (T_R(\tau ) \otimes T_L(\tau ))\;, \qquad \forall\ \sigma \ , \; 1\le \sigma \le r\;,\]
in such a way that:
\Bean
\item $T\RxL\{\sigma \}$ operates on the extended bilinear Hilbert subspace $H^+_a\{\sigma \}$
\[ T\RxL\{\sigma \}:\quad H^+_a\{\sigma \}\longrightarrow H^+_{ap}\{\sigma \}\]
sending it into the corresponding shifted  extended bilinear Hilbert subspace $H^+_{ap}\{\sigma \}$~.

\item $T\RxL\{\sigma \}\in \MM^a\RxL(H^+_a\{\sigma \})\;, \quad \forall\ \sigma $~,

where $\MM^a\RxL(H^+_a\{\sigma \})=\bigoplus_{\tau =1}^\sigma \MM^a\RxL (H^+_a(\tau ))$ is the direct sum of $\sigma $ subbialgebras of von Neumann.
\Ee
\vskip 11pt

\item As a consequence, a tower
\[ \MM^a\RxL(H^+_a\{1\}) \subset \cdots \subset
\MM^a\RxL(H^+_a\{\sigma \}) \subset \cdots \subset
\MM^a\RxL(H^+_a\{r\}) \]
of sums of von Neumann subbialgebras is generated.

\vskip 11pt

\item Similarly, a tower
\[ \MM^a\RxL(\Hs^+_a\{1\}) \subset \cdots \subset
\MM^a\RxL(\Hs^+_a\{\sigma \}) \subset \cdots \subset
\MM^a\RxL(\Hs^+_a\{r\}) \]
of sums of von Neumann subbialgebras on the diagonal bilinear Hilbert subspaces 
$\Hs^+_a\{\sigma \}$~, $\forall\ \sigma $~, can be introduced.
\Ei
\vskip 11pt

\subsection{Proposition}

{\em The discrete spectrum $\Sigma (T\RxL)$ of the (bi)operator $T\RxL\in \MM^a\RxL(H^+_a)$ is obtained by the set of isomorphisms:
\begin{align*}
i^a_{\{\sigma \}^D\RxL}: \quad
\MM^a\RxL(H^+_a\{\sigma \}) &\longrightarrow \MM^a\RxL(\Hs^+_a\{\sigma \})\\
T\RxL &\longrightarrow \Sigma (T\RxL)\;, \qquad 1\le \sigma \le r\;,\end{align*}
in such a way that to the set
\[ \lambda \RxL\{1\} \ , \; \cdots\ , \; 
\lambda \RxL\{\sigma \}\ , \; \cdots\ , \; 
\lambda \RxL\{r\}\;,\]
of eigenvalues of $\Sigma (T\RxL)$ corresponds to the set
\[ \widetilde M^{TS_+}_{ST_{L_R}}\{1\} \otimes_D \widetilde M^{TS_+}_{ST_{L}}\{1\} 
\ , \; \cdots\ , \; 
\widetilde M^{TS_+}_{ST_{L_R}}\{\sigma \} \otimes_D \widetilde M^{TS_+}_{ST_{L}}\{\sigma \} 
\ , \; \cdots\ , \; 
\widetilde M^{TS_+}_{ST_{L_R}}\{r\} \otimes_D \widetilde M^{TS_+}_{ST_{L}}\{r\} 
\]
of eigenbivectors which are (bi)states of the space-time field
$\widetilde M^{TS_+}_{ST_{L_R}} \otimes_D \widetilde M^{TS_+}_{ST_{L}}\approx \Hs^+_a\{r\}$~.
}

\bpr
\Bi
\item Referring to proposition 4.1.2, the biaction $T\RxL\{\sigma \}$ of the bioperator $T_R\otimes T_L$ (restricted to the partial sum $\{\sigma \}$ in the sense of section 4.7), on the extended bilinear Hilbert subspace $H^+_a\{\sigma \}$~:
\[T\RxL\{\sigma \} : \quad H^+_a\{\sigma \} \longrightarrow H^+_{ap}\{\sigma \}\]
sends $H^+_a\{\sigma \}$ into its shifted  equivalent $H^+_{ap}\{\sigma \}$ in such  way that:
\[T\RxL\{\sigma \}\in \MM^a\RxL(H^+_a\{\sigma \})=\bigoplus_{\sigma =1}^\tau 
\MM^a\RxL(H^+_a(\tau ))\;.\]
\vskip 11pt

\item The isomorphism $i^a_{\{\sigma \}^D\RxL}$ then corresponds to the map $H^+_{ap}\{\sigma \}\rightarrow \Hs^+_{ap}\{\sigma \}$ which:
\Bi
\item sends the shifted  extended bilinear Hilbert subspace $H^+_{ap}\{\sigma \}$ into its diagonal equivalent $\Hs^+_{ap}\{\sigma \}$~.
\item corresponds to the blowup isomorphism $S_L$ of proposition 3.19, applied to $H^+_{ap}\{\sigma \}\simeq \widetilde M^{TS_{p_+}}_{ST_{L_R}} \{\sigma \} \otimes 
\widetilde M^{TS_{p_+}}_{ST_{L}} \{\sigma \} $~, with the supplementary condition that the magnetic and electric fields
$ \widetilde M^{TS_{p_+}}_{ST_{L_R}} \{\sigma \} \otimes_{\rm magn}
\widetilde M^{TS_{p_+}}_{ST_{L}} \{\sigma \} $ and
$\widetilde M^{TS_{p_+}}_{ST_{L_R}} \{\sigma \} \otimes _{\rm elec}
\widetilde M^{TS_{p_+}}_{ST_{L}} \{\sigma \} $ be mapped onto the diagonal shifted  bilinear Hilbert subspace
$ \Hs^+_{ap}\{\sigma \} \simeq
\widetilde M^{TS_{p_+}}_{ST_{L_R}} \{\sigma \} \otimes _D
\widetilde M^{TS_{p_+}}_{ST_{L}} \{\sigma \} $~.
\Ei
\vskip 11pt

\item So, $\Hs^+_{ap}\{\sigma \} $ is generated and results from the map:
\[ T\RxDL\{\sigma \} : \quad \Hs^+_{a}\{\sigma \} \longrightarrow 
\Hs^+_{ap}\{\sigma \} \;, \quad T\RxDL\{\sigma \}\in \MM^a\RxL(\Hs^+_a\{\sigma \})\]
in such a way that $\lambda \RxL\{\sigma \}: T\RxDL\{\sigma \}\rightarrow \rit$ (or $\CC$~) is the eigenbivalue associated with $\{\sigma \}$ and corresponding to the (bi)generator of the respective Lie bialgebra introduced subsequently.
\vskip 11pt

\item $\widetilde M^{TS_{+}}_{ST_{L_R}} \{\sigma \} \otimes_D
\widetilde M^{TS_{+}}_{ST_{L}} \{\sigma \} \subseteq \Hs^+_a\{\sigma \}$ is then the $\sigma $-th eigenbivector, i.e. the {\bbf $\sigma $-th (bi)state of the vacuum space-time field\/} 
$\widetilde M^{TS_{+}}_{ST_{L_R}} \{\sigma \} \otimes_D
\widetilde M^{TS_{+}}_{ST_{L}} \{\sigma \} $~.

And its shifted  equivalent
$\widetilde M^{TS_{p_+}}_{ST_{L_R}} \{\sigma \} \otimes_D
\widetilde M^{TS_{p_+}}_{ST_{L}} \{\sigma \} \subseteq \Hs^+_{ap}\{\sigma \}$ is  the {\bbf $\sigma $-th (bi)state of the operator valued field\/} 
$\widetilde M^{TS_{p_+}}_{ST_{L_R}} \otimes_D
\widetilde M^{TS_{p_+}}_{ST_{L}} $~.\epr
\Ei
\vskip 11pt

\subsection{Deformations of states of the vacuum space-time operator valued field}

\Bi
\item Referring to section 4.1.6, (bi)projectors $P^{ap}\RxDL\{\sigma \}$ on the 
solvable shifted  diagonal bilinear Hilbert space $\Hs^+_{ap} \equiv \Hs^+_{ap}\{r\}$ 
can be introduced by the maps:
\begin{alignat*}{3}
P^{ap}\RxDL\{\sigma \} :\qquad\qquad\qquad
\Hs^+_{ap} 	&\longrightarrow  \Hs^+_{ap}\{\sigma \}\;, \qquad \forall\ \sigma \ , \; 1\le \sigma \le r\;, \\
 \widetilde M^{TS_{p_+}}_{ST_{L_R}} \otimes_D\widetilde M^{TS_{p_+}}_{ST_{L}} 
& \longrightarrow \widetilde M^{TS_{p_+}}_{ST_{L_R}} \{\sigma \} \otimes_D
\widetilde M^{TS_{p_+}}_{ST_{L}} \{\sigma \}\;.\end{alignat*}
Now, according to the chapter 1 of \cite{Pie4}, a projection acted by the map
$P^{ap}\RxDL\{\sigma \}$ on $\Hs^+_{ap}$ corresponds to an inverse deformation of a modular Galois representation studied by B. Mazur \cite{Maz2}.
\vskip 11pt

\item Indeed, a deformation $D^{\sigma \to r}\RxL$ of the $\sigma $-th bistate
$\widetilde M^{TS_{p_+}}_{ST_{L_R}} \{\sigma \} \otimes_D \widetilde M^{TS_{p_+}}_{ST_{L}} \{\sigma \}$ of the operator valued field
$\widetilde M^{TS_{p_+}}_{ST_{L_R}} \{r\} \otimes_D \widetilde M^{TS_{p_+}}_{ST_{L}} \{r\}$ corresponds to the injective mapping:
\[ D^{\sigma \to r}\RxL: \quad
\widetilde M^{TS_{p_+}}_{ST_{L_R}} \{\sigma \} \otimes_D \widetilde M^{TS_{p_+}}_{ST_{L}} \{\sigma \}\longrightarrow 
\widetilde M^{TS_{p_+}}_{ST_{L_R}} \{r\} \otimes_D \widetilde M^{TS_{p_+}}_{ST_{L}} \{r\}\]
such that $P^{ab}\RxDL\{\sigma \}=(D^{\sigma \to r}\RxL)^{-1}$~.

This deformation is associated with the exact sequence
\begin{multline*}
\qquad 0 \longrightarrow 
\widetilde M^{TS_{p_+}}_{ST_{L_R}} \{1\} \otimes_D \widetilde M^{TS_{p_+}}_{ST_{L}} \{1\}
\longrightarrow 
\widetilde M^{TS_{p_+}}_{ST_{L_R}} \{r\} \otimes_D \widetilde M^{TS_{p_+}}_{ST_{L}} \{r\}\\
\longrightarrow 
\widetilde M^{TS_{p_+}}_{ST_{L_R}} \{\sigma \} \otimes_D \widetilde M^{TS_{p_+}}_{ST_{L}} \{\sigma \}\longrightarrow 0\qquad\end{multline*}
whose kernel is
$\widetilde M^{TS_{p_+}}_{ST_{L_R}} \{1\} \otimes_D \widetilde M^{TS_{p_+}}_{ST_{L}} \{1\}$~.
\Ei
\vskip 11pt

\subsection{Proposition: quantization rules}

{\em Let $\rho $ be an index $\in\NN$ running on $r-\sigma $~.

\Bi
\item Then, the deformation
\[ D^{\sigma \to r}\RxL: \quad
\widetilde M^{TS_{p_+}}_{ST_{L_R}} \{\sigma \} \otimes_D \widetilde M^{TS_{p_+}}_{ST_{L}} \{\sigma \}\longrightarrow 
\widetilde M^{TS_{p_+}}_{ST_{L_R}} \{r\} \otimes_D \widetilde M^{TS_{p_+}}_{ST_{L}} \{r\}\]
corresponds to a quantization rule consisting in adding $\sum\limits_\rho m^{(\rho )}$ closed bistrings to the $\sigma $-th (bi)state of the operator valued string field
$
\widetilde M^{TS_{p_+}}_{ST_{L_R}} \{r\} \otimes_D \widetilde M^{TS_{p_+}}_{ST_{L}} \{r\} $~.

\item And, the inverse deformation $(D^{\sigma \to r}\RxL)^{-1}\equiv
P^{ap}\RxDL\{\sigma \}$ corresponds to the quantization rule consisting in extracting
$\sum\limits_\rho  m^{(\rho )}$ closed bistrings from the operator valued string field
$ \widetilde M^{TS_{p_+}}_{ST_{L_R}} \{r\} \otimes_D \widetilde M^{TS_{p_+}}_{ST_{L}} \{r\} $~.
\Ei}
\vskip 11pt

\bpr
\Bi
\item In fact, the inverse transformation $(D^{\sigma \to r}\RxL)^{-1}$ is the map:
\begin{multline*}
(D^{\sigma \to r}\RxL)^{-1}: \quad
\widetilde M^{TS_{p_+}}_{ST_{L_R}} \{r\} \otimes_D \widetilde M^{TS_{p_+}}_{ST_{L}} \{r\} \\
\longrightarrow \widetilde M^{TS_{p_+}}_{ST_{L_R}} \{\sigma \} \otimes_D \widetilde M^{TS_{p_+}}_{ST_{L}} \{\sigma \} \bigoplus_\rho \bigoplus_{m_\rho }
(\widetilde M^{TS_{p_+}}_{ST_{L_R}} (\rho ,m_\rho )\otimes_D \widetilde M^{TS_{p_+}}_{ST_{L}} (\rho ,m_\rho ))
\end{multline*}
generating free shifted closed bistrings
$\widetilde M^{TS_{p_+}}_{ST_{L_R}}(\rho ,m_\rho )\otimes_D \widetilde M^{TS_{p_+}}_{ST_{L}} (\rho ,m_\rho )$ at $\rho $ biquanta, where $m^{(\rho )}=\sup(m_\rho )+1$ is the multiplicity of the $\rho $-th section of the shifted  bisemisheaf
$\widetilde M^{TS_{p_+}}_{ST_{L_R}}  \otimes_D \widetilde M^{TS_{p_+}}_{ST_{L}} $ according to  proposition 4.1.2.
\vskip 11pt

\item $(D^{\sigma \to r}\RxL)^{-1}$ corresponds to an endomorphism based on Galois antiautomorphisms \cite{Pie6} removing free shifted closed bistrings as described in \cite{Pie4}.\epr
\Ei
\vskip 11pt

\section{Creations and annihilations of mass string fields}

\subsection{Fusion at the Planck scale of GR with QFT}

In chapter 3 and 4.1,  it was seen how (brane and) string fields of the vacuum internal structure of (bisemi)fermions as well as their states could be generated algebraically.

As indicated in section 2.7, strong fluctuations occur on these vacuum string fields $ \widetilde M^{TS}_{ST_{L_R}} \otimes_D \widetilde M^{TS}_{ST_{L}}$ because they have a spatial extension of the order of the Planck length.

These strong fluctuations generate singularities on the sections (or strings) of these vacuum space-time fields in such a way that, if these singularities are degenerate, they are able to produce, by versal deformations and blowups of these, two new covering space-time fields of which the most external can be interpreted as mass fields of these (bisemi)fermions.

By this way, general relativity can be connected at the microscopic level to quantum field theory as developed in section 2.6.  And, the set of vacuum string fields
$\widetilde M^{TS}_{ST_{L_R}}\otimes_D \widetilde M^{TS}_{ST_{L}}$ could correspond to the dark energy which develops in this context a dynamical aspect \cite{Pie2} since it is able to create mass fields of elementary particles.

The aim of the next following sections consists in showing how mass strings can be created from the vacuum string field $\widetilde M^{TS}_{ST_{L_R}}\otimes_D\widetilde M^{TS}_{ST_{L}}$ by the blowup of the versal deformations.
\vskip 11pt

\subsection{Versal deformations}

We refer to \cite{Pie4} for the technical aspects of the versal deformation and of its blowup, called spreading-out.

\Bi
\item Let $\widetilde M^{S}_{ST_{L}}$ (resp. $\widetilde M^{S}_{ST_{R}}$~) 
denote the \lr space semisheaf of the vacuum string field.  The set 
$\{\widetilde M^{S}_{v_{\mu ,m_\mu }}\} ^q_{\mu =1,m_\mu } $ 
(resp. $\{\widetilde M^{S}_{\o v_{\mu ,m_\mu }}\}^q_{\mu =1,m_\mu } $~) of its $q$ sections, having multiplicities $m^\mu =\sup(m_\mu )+1$~, are one-dimensional $\CC$-valued differentiable functions over the respective conjugacy class representatives of $T_2(L_v)$ (resp. $T^t_2(L_{\o v}$~) according to section 3.6: these sections are strings.
\vskip 11pt

\item It is assumed that, under a strong external perturbation, a degenerate singularity of multiplicity 3 is generated on each section
$\widetilde M^{S}_{v_{\mu ,m_\mu }}$ (resp. $\widetilde M^{S}_{\o v_{\mu ,m_\mu }}$~) 
of
$\widetilde M^{S}_{ST_{L}}$ (resp. $\widetilde M^{S}_{ST_{R}}$~).
\vskip 11pt

\item Then, a versal deformation of $\widetilde M^{S}_{ST_{L}}$ (resp. $\widetilde M^{S}_{ST_{R}}$~) will be given by the fiber bundle:
\begin{align*}
D_{S_L} : \quad & \widetilde M^{S}_{ST_{L}}  \times \theta _{S_L} 
\longrightarrow \widetilde M^{S}_{ST_{L}}\\
\text{(resp.} \quad 
D_{S_L} : \quad & \widetilde M^{S}_{ST_{R}}  \times \theta _{S_R} 
\longrightarrow \widetilde M^{S}_{ST_{R}}\ )\end{align*}
in such a way that the fiber $\theta _{S_L}=\{ \theta ^1(\omega ^1_L),
\theta ^2(\omega ^2_L),\theta ^3(\omega ^3_L)\}$ (resp.
$\theta _{S_R}=\{ \theta ^1(\omega ^1_R),
\theta ^2(\omega ^2_R),\theta ^3(\omega ^3_R)\}$~) is composed of three sheaves of the base $S_L$ (resp. $S_R$~) of the versal deformation, the $u_i\omega ^i_R$ (resp. $u_i\omega ^i_L$~), $1\le i\le 3$~, $u_i\in\rit$~, being the monomials of the rest polynomials (of the quotient algebra)
\begin{align*}
R_{v_{\mu ,m_\mu }} 
&= \sum^3_{i=1} u_i(v_{\mu ,m_\mu} )\omega ^i_L(v_{\mu ,m_\mu })\\
\text{(resp.} \quad 
R_{\o v_{\mu ,m_\mu }} &= \sum^3_{i=1} u_i(\o v_{\mu ,m_\mu} )\omega ^i_R(\o v_{\mu ,m_\mu })\ )\end{align*}
of the versal unfoldings of the singularities on the sections 
$\widetilde M^{S}_{v_{\mu ,m_\mu }}$ (resp. $\widetilde M^{S}_{\o v_{\mu ,m_\mu }}$~) 
following the preparation theorem.
\vskip 11pt

\item The fiber $\theta _{S_L}$ (resp. $\theta _{S_R}$~) is of algebraic 
nature in the sense that each function $\omega ^i_L(v_{\mu ,m_\mu} )$ (resp.
$\omega ^i_R(\o v_{\mu ,m_\mu })$~) is defined over $\tau _{ i,\mu }$ quanta, $\tau _{i,\mu }\in \NN$~, and is thus characterized by a rank or degree equal to 
$\tau _{ i,\mu }\centerdot N$~.

The versal unfolding of a singularity then consists in ``pumping'' external free quanta which are projected in the neighborhood of the singularity in order to stabilize it.
\Ei\vskip 11pt

\subsection{Blowup of the versal deformation}

\Bi
\item The blowup of the versal deformation of the semisheaf 
$\widetilde M^{S}_{ST_{L}}$ (resp. $\widetilde M^{S}_{ST_{R}}$~) is realized by the spreading-out isomorphism:
\[ SOT_L = (\tau _{v_{\omega _L}}\circ \pi _{s_L}) \qquad \text{(resp.} \quad
SOT_R= (\tau _{v_{\omega _R}}\circ \pi _{s_R})\ )\]
where
\Bi
\item $\begin{array}[t]{rll}
\pi _{s_L} : \quad & \widetilde M^{S}_{ST_{L}}\times \theta _{S_L} &\longrightarrow 
\widetilde M^{S}_{ST_{L}}\cup \theta _{S_L}\\
\text{(resp.} \quad 
\pi _{s_R} : \quad & \widetilde M^{S}_{ST_{R}} \times \theta _{S_R} &\longrightarrow 
\widetilde M^{S}_{ST_{R}}\cup \theta _{S_R}\ )\end{array}$

is an endomorphism disconnecting the three base sheaves $\theta _{S_L}$ 
(resp. $\theta _{S_R}$~) from $\widetilde M^{S}_{ST_{L}}$ (resp. $
\widetilde M^{S}_{ST_{R}}$~).

\item $\tau _{v_{\omega L}}$ (resp. $\tau _{v_{\omega R}}$~) is the projective map:
\begin{align*}
\tau _{v_{\omega L}}: \quad &\TAN(\theta _{S_L})\longrightarrow \theta _{S_L}\\
\text{(resp.} \quad 
\tau _{v_{\omega R}}: \quad &\TAN(\theta _{S_R})\longrightarrow \theta _{S_R}\ )\end{align*}
of the vertical tangent bundle $T_{{\rm v}_{\omega L}}$ (resp. $T_{{\rm v}_{\omega R}}$~) sending $\theta _{S_L}$ (resp. $\theta _{S_R}$~) in the total tangent space $\TAN(\theta _{S_L})$ (resp. $\TAN(\theta _{S_R})$~).
\Ei
\vskip 11pt

\item The spreading-out isomorphism then projects the three functions $\omega ^i_L(v_{\mu ,m_\mu })$ (resp. $\omega ^i_R(\o v_{\mu ,m_\mu })$~) above each section
$\widetilde M^{S}_{v_{\mu ,m_\mu }}$ (resp. $\widetilde M^{S}_{\o v_{\mu ,m_\mu }}$~) in the vertical tangent space in such a way that these three functions
$\{\omega ^i_L(v_{\mu ,m_\mu }) \}^3_{i=1}$ (resp. $\{\omega ^i_R(\o v_{\mu ,m_\mu })\}^3_{i=1}$~) cover 
$\widetilde M^{S}_{v_{\mu ,m_\mu }}$ (resp. $\widetilde M^{S}_{\o v_{\mu ,m_\mu }}$~).
\vskip 11pt

\item After that, these three functions are glued together in a compact way: they then generate sections
$\widetilde M^{S}_{MG_{v_{\mu ,m_\mu }}}$ (resp. $\widetilde M^{S}_{MG_{\o v_{\mu ,m_\mu }}}$~) of a semisheaf 
$\widetilde M^{S}_{MG_L}$ (resp. $\widetilde M^{S}_{MG_R}$~) (called middle ground) which cover the internal vacuum smisheaf
$\widetilde M^{S}_{ST_L}$ (resp. $\widetilde M^{S}_{ST_R}$~).
\vskip 11pt

\item If the numbers of quanta on the sections
$\widetilde M^{S}_{MG_{v_{\mu ,m_\mu }}}$ (resp. $\widetilde M^{S}_{MG_{\o v_{\mu ,m_\mu }}}$~) of $\widetilde M^{S}_{MG_L}$ (resp. $\widetilde M^{S}_{MG_R}$~) are equal to the numbers of quanta on the sections 
$\widetilde M^{S}_{v_{\mu ,m_\mu }}$ (resp. $\widetilde M^{S}_{\o v_{\mu ,m_\mu }}$~), rewritten according to
$\widetilde M^{S}_{ST_{v_{\mu ,m_\mu }}}$ (resp. $\widetilde M^{S}_{ST_{\o v_{\mu ,m_\mu }}}$~) of $\widetilde M^{S}_{ST_L}$ (resp. $\widetilde M^{S}_{ST_R}$~), then these sections $\widetilde M^{S}_{MG_{v_{\mu ,m_\mu }}}$ (resp. $\widetilde M^{S}_{MG_{\o v_{\mu ,m_\mu }}}$~) are open strings covering the closed strings
$\widetilde M^{S}_{ST_{v_{\mu ,m_\mu }}}$ (resp. $\widetilde M^{S}_{ST_{\o v_{\mu ,m_\mu }}}$~) of $\widetilde M^{S}_{ST_L}$ (resp. $\widetilde M^{S}_{ST_R}$~).
\Ei
\vskip 11pt

\subsection[Generation of mass semisheaves $\widetilde M^{S}_{M_L}$ and $\widetilde M^{S}_{M_R}$]{{\boldmath Generation of mass semisheaves $\widetilde M^{S}_{M_L}$ and $\widetilde M^{S}_{M_R}$}}

\Bi
\item As the degenerate singularities on the sections of
$\widetilde M^{S}_{ST_L}$ (resp. $\widetilde M^{S}_{ST_R}$~) are of multiplicity 3, 
the functions $\omega ^i_L(v_{\mu ,m_\mu })$ (resp. 
$\omega ^i_R(\o v_{\mu ,m_\mu })$~)   of the quotient algebra of the versal deformation of
$\widetilde M^{S}_{ST_L}$ (resp. $\widetilde M^{S}_{ST_R}$~) can have degenerate 
singularities of multiplicity one.

So, a versal deformation of the semisheaf
$\widetilde M^{S}_{MG_L}$ (resp. $\widetilde M^{S}_{MG_R}$~) and a blowup of it can be envisaged as for $\widetilde M^{S}_{ST_L}$ (resp. $\widetilde M^{S}_{ST_R}$~).
\vskip 11pt

\item As a consequence, a mass semisheaf $\widetilde M^{S}_{M_L}$ (resp. $\widetilde M^{S}_{M_R}$~) can be generated algebraically from the middle-ground semisheaf
$\widetilde M^{S}_{MG_L}$ (resp. $\widetilde M^{S}_{MG_R}$~) according to the composition of maps:
\begin{align*}
SOT_L^{(MG)}\circ D_{S_L}^{(MG)}: \quad 
\widetilde M^{S}_{MG_L} &\longrightarrow \widetilde M^{S}_{MG_L} \cup
\widetilde M^{S}_{M_L} \\
\text{(resp.} \quad 
SOT_R^{(MG)}\circ D_R^{(MG)}: \quad 
\widetilde M^{S}_{MG_R} &\longrightarrow \widetilde M^{S}_{MG_R} \cup
\widetilde M^{S}_{M_R} \ )\end{align*}
in such a way that the sections 
$\widetilde M^{S}_{M_{v_{\mu ,m_\mu }}}$ (resp. 
$\widetilde M^{S}_{M_{\o v_{\mu ,m_\mu }}}$~) of $\widetilde M^{S}_{M_L}$ (resp. $\widetilde M^{S}_{M_R}$~), which cover the corresponding sections of
$\widetilde M^{S}_{MG_L}$ (resp. $\widetilde M^{S}_{MG_R}$~) and of
$\widetilde M^{S}_{ST_L}$ (resp. $\widetilde M^{S}_{ST_R}$~), are open strings if they have the same numbers of quanta as the sections of
$\widetilde M^{S}_{ST_L}$ (resp. $\widetilde M^{S}_{ST_R}$~).
\Ei
\vskip 11pt

\subsection{Generation of middle ground and mass fields}

\Bi\item So, by versal deformation and blowup of it, the middle ground and mass semisheaves of space 
$\widetilde M^{S}_{MG_L}$ (resp. $\widetilde M^{S}_{MG_R}$~) and 
$\widetilde M^{S}_{M_L}$ (resp. $\widetilde M^{S}_{M_R}$~) can be generated algebraically from the internal vacuum semisheaf
$\widetilde M^{S}_{ST_L}$ (resp. $\widetilde M^{S}_{ST_R}$~) so that one has the following embedding:
\begin{align*}
\widetilde M^{S}_{ST_L} &\subset \widetilde M^{S}_{MG_L} \subset \widetilde M^{S}_{M_L}\\
\text{(resp.} \quad 
\widetilde M^{S}_{ST_R} &\subset \widetilde M^{S}_{MG_R} \subset \widetilde M^{S}_{M_R}\ ).\end{align*}
\vskip 11pt

\item Similar developments can be envisaged to generate the middle ground and mass semisheaves of time 
$\widetilde M^{T}_{MG_L}$ (resp. $\widetilde M^{T}_{MG_R}$~) and 
$\widetilde M^{T}_{M_L}$ (resp. $\widetilde M^{T}_{M_R}$~) from the internal vacuum semisheaf of time
$\widetilde M^{T}_{ST_L}$ (resp. $\widetilde M^{T}_{ST_R}$~) leading to the embedding:
\begin{align*}
\widetilde M^{T}_{ST_L} &\subset \widetilde M^{T}_{MG_L} \subset \widetilde M^{T}_{M_L}\\
\text{(resp.} \quad 
\widetilde M^{T}_{ST_R} &\subset \widetilde M^{T}_{MG_R} \subset \widetilde M^{T}_{M_R}\ ).\end{align*}
\vskip 11pt

\item And, if it was the case, the corresponding semisheaves of space could be generated from their corresponding time semisheaves by $ (\gamma _{t\to r}\circ E)$
morphisms where $E$ is an endomorphism based on Galois antiautomorphisms as developed in chapter 1 of \cite{Pie4}.
\vskip 11pt

\item By this way, middle ground and mass fields of space-time
$\widetilde M^{TS}_{MG_R}  \otimes_D\widetilde M^{TS}_{MG_L}$ and 
$\widetilde M^{TS}_{M_R}  \otimes_D\widetilde M^{TS}_{M_L}$ are produced from the vacuum most internal field
$\widetilde M^{TS}_{ST_R}  \otimes_D\widetilde M^{TS}_{ST_L}$ of space-time, leading to the embedding:
\[
\widetilde M^{TS}_{ST_R}  \otimes_D\widetilde M^{TS}_{ST_L} 
\quad \subset\quad 
\widetilde M^{TS}_{MG_R}  \otimes_D\widetilde M^{TS}_{MG_L} 
\quad \subset\quad 
\widetilde M^{TS}_{M_R}  \otimes_D\widetilde M^{TS}_{M_L} \;.\]
Indeed:\quad $\widetilde M^{TS}_{M_L}  
= \widetilde M^{T}_{M_L}  \bigoplus \widetilde M^{S}_{M_L}$~, and so on.
\vskip 11pt

\item It must be noticed that the left and right middle ground and mass semisheaves are produced symmetrically since it is assumed that:
\Bean
\item they are centered on the emergence point (local origin) of the elementary (bisemi)fermion.
\item the perturbations, generating singularities, are identical locally around the emergence point in the upper and lower half spaces.
\Ee
\Ei
\vskip 11pt

\subsection{Proposition}

{\em
Let $\widetilde M^{TS}_{ST-MG_R}  \otimes_D\widetilde M^{TS}_{ST-MG_L} 
\equiv  (\widetilde M^{TS}_{ST_R}  \bigoplus\widetilde M^{TS}_{MG_R} )\otimes_D
(\widetilde M^{TS}_{ST_L}  \bigoplus\widetilde M^{TS}_{MG_L} ) $ denote the space-time (~$ST$~) and middle-ground (~$MG$~) fields of the vacuum of an elementary bisemifermion.

Then, by versal deformation and blowup of it, the middle-ground field
$\widetilde M^{TS}_{MG_R}  \otimes_D\widetilde M^{TS}_{MG_L} $ can {\bf create\/}, section after section, {\bbf the open bistrings
$ \widetilde M^{TS}_{M_{\o v_{\sigma ,m_\sigma }}}  \otimes_D 
\widetilde M^{TS}_{M_{v_{\sigma ,m_\sigma }}}  $ of the mass field\/}
$\widetilde M^{TS}_{M_R}  \otimes_D\widetilde M^{TS}_{M_L} $ according to:
\begin{multline*}
SOT^{(MG)}\RxL \circ D^{(MG)}_{S\RxL} : \qquad
\widetilde M^{TS}_{ST-MG_R}  \otimes_D\widetilde M^{TS}_{ST-MG_L} \\
\longrightarrow 
\widetilde M^{TS}_{ST-MG_R}  \otimes_D\widetilde M^{TS}_{ST-MG_L} 
\cup \{
 \widetilde M^{TS}_{M_{\o v_ {\sigma ,m_\sigma} }}  \otimes_D 
\widetilde M^{TS}_{M_{v_{\sigma ,m_\sigma }}} 
 \}^r_{\sigma =1,m_\sigma }\end{multline*}
in such a way that the mass open bistrings
$
 \widetilde M^{TS}_{M_{\o v_{\sigma ,m_\sigma }}}  \otimes_D 
\widetilde M^{TS}_{M_{v_{\sigma ,m_\sigma }}  }
\subset
 \widetilde M^{TS}_{M_{\o \omega _{\sigma  }}}  \otimes_D 
\widetilde M^{TS}_{M_{\omega _{\sigma}}  }$~, included into the corresponding mass openbibranes
$ \widetilde M^{TS}_{M_{\o \omega _{\sigma  }}}  \otimes_D 
\widetilde M^{TS}_{M_{\omega _{\sigma}}  }$~, cover the corresponding ``~$ST$~'' and ``~$MG$~'' bistrings
$\widetilde M^{TS}_{ST_{\o v_{\sigma ,m_\sigma }}}  \otimes_D 
\widetilde M^{TS}_{ST_{v_{\sigma ,m_\sigma }} } $ and
$\widetilde M^{TS}_{MG_{\o v_{\sigma ,m_\sigma }}}  \otimes_D 
\widetilde M^{TS}_{MG_{v_{\sigma ,m_\sigma }} } $~, included into their corresponding bibranes (see proposition 3.7).
}
\vskip 11pt

\bpr Referring to section 4.2.4, we see that the versal deformation $D^{(MG)}_{S\RxL}$
of the middleground field 
$\widetilde M^{TS}_{MG_R } \otimes_D \widetilde M^{TS}_{MG_L} $~, following by its blowup $SOT ^{(MG)}\RxL$~, generates the mass field
$
\widetilde M^{TS}_{M_R } \otimes_D \widetilde M^{TS}_{M_L} $ section after section.\epr
\vskip 11pt

\subsection{Corollary}
{\em
Let $\widetilde M^{TS}_{ST-MG-M_R}  \otimes_D\widetilde M^{TS}_{ST-MG-M_L} 
$ denote the vacuum fields
$\widetilde M^{TS}_{ST_R}  \otimes_D\widetilde M^{TS}_{ST_L} $
and
$\widetilde M^{TS}_{MG_R}  \otimes_D\widetilde M^{TS}_{MG_L}$
of a bisemifermion covered by its mass field
 $\widetilde M^{TS}_{M_R}  \otimes_D\widetilde M^{TS}_{M_L} $~.

Then, a set
$\{  
\widetilde M^{TS}_{M_{\o v_{\sigma,m_\sigma   }}}  \otimes_D 
\widetilde M^{TS}_{M_{v_{\sigma,m_\sigma }}  }\}_{m_\sigma }$ of
$m^{(\sigma )}$ (~$=\sup (m_\sigma)+1 $~) {\bf mass open bistrings\/}, characterized by $\sigma $ biquanta, {\bf are annihilated\/} if they become free, i.e. are disconnected from the mass field
$ \widetilde M^{TS}_{M_R}  \otimes_D 
\widetilde M^{TS}_{M_L  } $~.
}
\vskip 11pt

\bpr This is realized by considering the smooth endomorphism:
\begin{multline*}
E_{M\RxL} : \qquad
\widetilde M^{TS}_{ST-MG-M_R}  \otimes_D\widetilde M^{TS}_{ST-MG-M_L} \\
\longrightarrow 
\widetilde M^{TS}_{ST-MG-(M\setminus\sigma )_R}  \otimes_D
\widetilde M^{TS}_{ST-MG-(M\setminus\sigma )_L} 
\bigoplus_{m_\sigma } \{
 \widetilde M^{TS}_{M_{\o v_ {\sigma ,m_\sigma} }}  \otimes_D 
\widetilde M^{TS}_{M_{v_{\sigma ,m_\sigma }}} 
 \}_{m_\sigma }\end{multline*}
with the evident notation $(M\setminus \sigma )$~.\epr
\vskip 11pt

\subsection{Proposition: quantum jumps}

{\em \Bean
\item A set 
$\{ \widetilde M^{TS}_{ST-MG-M_{\o v_{\sigma ,m_\sigma }}}  \otimes_D 
\widetilde M^{TS}_{ST-MG-M_{v_{\sigma ,m_\sigma }}} \}_{m_\sigma }$ of $m^{(\sigma )}$ 
bistrings, i.e. (bisemi)photons, on the ``~$ST$~'', ``~$MG$~'' and ``~$M$~'' 
fields are emitted from
$\widetilde M^{TS}_{ST-MG-M_R}  \otimes_D 
\widetilde M^{TS}_{ST-MG-M_L}$ 
if they become free, i.e. are disconnected from these fields.

\item A set 
$\{ \widetilde M^{TS}_{ST-MG-M_{\o v_{\sigma ,m_\sigma }}}  \otimes_D 
\widetilde M^{TS}_{ST-MG-M_{v_{\sigma ,m_\sigma }}} \}_{m_\sigma }$ of $m^{(\sigma )}$ free bistrings, i.e. (bisemi)photons, can be absorbed by the fields
$\widetilde M^{TS}_{ST-MG-M_R }  \otimes_D 
\widetilde M^{TS}_{ST-MG-M_L }$ if they become bisections of these bisemisheaves.
\Ee}
\vskip 11pt

\bpr
\Bean
\item The set 
of $m^{(\sigma )}$ bistrings  on the ``~$ST$~'', ``~$MG$~'' and ``~$M$~'' fields 
are emitted from
$\widetilde M^{TS}_{ST-MG-M_R }  \otimes_D 
\widetilde M^{TS}_{ST-MG-M_L }$ by considering the smooth endomorphism:
\begin{multline*}
E_{ST-MG-M\RxL} : \qquad
(\widetilde M^{TS}_{ST-MG-M_R}  \otimes_D\widetilde M^{TS}_{ST-MG-M_L}) \\
\longrightarrow 
(\widetilde M^{TS}_{ST-MG-M\setminus(\sigma )_R}  \otimes_D
\widetilde M^{TS}_{ST-MG-M\setminus(\sigma )_L} )
\bigoplus_{m_\sigma } \{
 \widetilde M^{TS}_{ST-MG-M_{\o v_ {\sigma ,m_\sigma} }}  \otimes_D 
\widetilde M^{TS}_{ST-MG-M_{v_{\sigma ,m_\sigma }}} 
 \}_{m_\sigma }\end{multline*}
simultaneously on the three fields
$
\widetilde M^{TS}_{ST-MG-M_R}  \otimes_D\widetilde M^{TS}_{ST-MG-M_L} $~.
\vskip 11pt

\item The same set 
of $m^{(\sigma )}$ bistrings  is absorbed by the three fields:
$\widetilde M^{TS}_{ST-MG-M_R }  \otimes_D 
\widetilde M^{TS}_{ST-MG-M_L }$ if we consider the inverse map
$E^{-1}_{ST-MG-M\RxL}$ introduced in a).\epr
\Ee\vskip 11pt

\section{Interacting fields of interacting bisemifermions}

\subsection{The importance of the twofold nature of the microscopic reality}

\Bi
\item Chapter 3 and 4 until now have dealt with the algebraic generation of the three embedded diagonal fields 
``~$ST$~'', ``~$MG$~'' and ``~$M$~'' constituting the central internal structure of a bisemifermion, without taking explicitly into account the electric (internal) field (i.e. the electric charge) and the internal magnetic field (i.e. the magnetic moment) except at the end of chapter 3.
\vskip 11pt

\item It appears thus that the internal structure of a bisemifermion is very complex, especially if we consider that off-diagonal fields of interaction exist between the three central diagonal fields
``~$ST$~'', ``~$MG$~'' and ``~$M$~'', as developed at the beginning of chapter 3 in \cite{Pie4}.
\vskip 11pt

\item The fact of considering that the nature at the microscopic scale is twofold allowed us to introduce the diagonal fields of elementary (bisemi)fermions and the off-diagonal magnetic and electric fields.

But, the twofold nature of reality is of crucial importance when the problem of interactions between (bisemi)particles is envisaged, as it will be done succinctly in the following sections.
\Ei
\vskip 11pt

\subsection{Non-orthogonal reducible modular representation space}

\Bi
\item As developed at the beginning of chapter 5 of \cite{Pie4}, the time or space string field(s)
``~$ST$~'', ``~$MG$~'' and ``~$M$~'' of a set of $M$ interacting bisemifermions is given by the completely reducible modular representation space
$\Repsp( \GL_{2M}(L_{\o v_+}\times L_{v_+}))$ of the bilinear general semigroup
$\GL_{2M}(L_{\o v_+}\times L_{v_+})$ (see section 3.2.e)).
\vskip 11pt

\item Given the partition $2M=2_1+2_2+\cdots+2_i+\cdots+2_M$ of $2M$~, the completely reducible modular bilinear {\bf non orthogonal representation space\/}
$\Repsp( \GL_{2M_{(i\neq j)}}(L_{\o v_+}\times L_{v_+}))$ decomposes into \cite{Pie5}:
\begin{multline*}
\Repsp( \GL_{2M_{(i\neq j)}}(L_{\o v_+}\times L_{v_+}))\\
= \bplus^M_{i=1}
\Repsp( \GL_{2_{i}}(L_{\o v_+}\times L_{v_+}))
\bplus^M_{i\neq j=1}
\Repsp( T^t_{2_i}(L_{\o v_+})\times T_{2j}(L_{v_+}))
\end{multline*}
while the corresponding {\bf orthogonal\/} representation space is given by:
\begin{multline*}
\Repsp( \GL_{2M_{(i)}}(L_{\o v_+}\times L_{v_+}))\\
= \boxplus^M_{i=1}
\Repsp( \GL_{2_{i}}(L_{\o v_+}\times L_{v_+}))
\subset
\Repsp( \GL_{2M_{i\neq j}}(L_{\o v_+}\times L_{v_+}))\;.
\end{multline*}
\vskip 11pt

\item So, the fact of considering bilinear algebraic semigroups allows to take into account off-diagonal modular representation spaces which are responsible for the generation of interacting fields between bisemiparticles as it will be seen in the next section.
\Ei
\vskip 11pt

\subsection{Gravito-electro-magnetic fields of interaction}

\Bi
\item Assume that
$\Repsp(
\GL_{2_{i}}(L_{\o v_+}\times L_{v_+}))$ is the string mass field of space
$\widetilde M^S_{M_{R_i}}\otimes_{(D)} \widetilde M^S_{M_{L_i}}$ of the $i$-th considered bisemifermion as envisaged previously.  Then, for a set of $M$ (interacting) bisemifermions, the string mass fields of space will be given by:
\[\Repsp(
\GL_{2M_{(i\neq j)}}(L_{\o v_+}\times L_{v_+}))
= \bigoplus^M_{i=1}
(\widetilde M^S_{M_{R_i}}\otimes_{(D)} \widetilde M^S_{M_{L_i}} )
\bigoplus^M_{i\neq j}
(\widetilde M^S_{M_{R_i}}\otimes_{(D)} \widetilde M^S_{M_{L_j}} )\]
where the
$(\widetilde M^S_{M_{R_i}}\otimes_{(D)} \widetilde M^S_{M_{L_j}} )$ are interacting mass fields of space which are gravitational and magnetic fields as proved in \cite{Pie4}.
\vskip 11pt

\item If these $M$ bisemifermions are free, then their string mass fields of space reduce to:
\[\Repsp(
\GL_{2M_{(i)}}(L_{\o v_+}\times L_{v_+}))
= \bigoplus^M_{i=1}
(\widetilde M^S_{M_{R_i}}\otimes_{(D)} \widetilde M^S_{M_{L_i}} )\;,\]
i.e. to their internal mass fields of space.
\vskip 11pt

\item If the complete internal structure of the $M$ bisemifermions is given by the fields $
\widetilde M^{TS}_{ST-MG-M_{R_i}}\otimes_{(D)} \widetilde M^{TS}_{ST-MG-M_{L_i}} )
$~, $1\le i\le M$~, as envisaged in section 4.2, then a set of gravito-electro-magnetic fields of interaction are generated between the
``~$ST$~'', ``~$MG$~'' and ``~$M$~'' internal fields of these bisemifermions as developed in chapter 5 in \cite{Pie4}.
\Ei
\vskip 11pt

\subsection{Bosonic character of the fields of the bisemifermions}

\Bi
\item
In AQT, the state(s) of the field(s) (for example, space field of mass) of a set of $M$ (free) bisemifermions can be constructed as the direct sum(s) of the state(s) of the $M$ individual space fields of mass according to sections 4.3.2 and 4.3.3.
\vskip 11pt

\item This contrasts with the treatment envisaged in QFT for the state (of the field) of a set of $M$ free fermions which is given as an antisymmetric superposition of the product of the individual states in order to obey the Pauli exclusion principle.
\vskip 11pt

\item As a consequence, the field of a set of $M$ (free) bisemifermions does not 
behave in AQT like a fermionic field of QFT but as a bosonic field, 
{\bf the fermionic character being given by the off-diagonal electric fields\/} of interaction, which corresponds to the electric charges at the individual fermionic levels.
\vskip 11pt

\item Indeed, QFT only works with the linear mass field (and, the not well defined vacuum field) of fermions while AQT has introduced time and space fields of bilinear type, which allows to encircle the fermionic character differently and more precisely as it was envisaged in QFT.
\Ei

\end{document}